\documentclass[twocolumn,tighten]{aastex631}
\graphicspath{{./}{figures/}}

   \newcommand{\ram}{$^{\mbox{\scriptsize m}}$}
   \newcommand{\ras}{$^{\mbox{\scriptsize s}}$}
   \newcommand{\decd}{$^{\circ}$}
   \newcommand{\decm}{$'$}

   \newcommand{\beam}{$\theta_{\mbox{\scriptsize maj}}\times\theta_{\mbox{\scriptsize min}}$}
   
   \newcommand{\ujyperbeam}{$\mu$Jy\,beam$^{-1}$}

\usepackage[none]{hyphenat}
\usepackage{array}
\newcolumntype{+}{>{\global\let\currentrowstyle\relax}}
\newcolumntype{^}{>{\currentrowstyle}}

\newcommand\clearrow{\global\let\rowmac\relax}
\clearrow

\shorttitle{Radio images on PDS~70}
\shortauthors{Liu et al.}

\begin{document}

\title{First JVLA Radio Observation on PDS~70}

\correspondingauthor{Hauyu Baobab Liu}
\email{hyliu.nsysu@g-mail.nsysu.edu.tw}

\author[0000-0003-2300-2626]{Hauyu Baobab Liu}
\affiliation{Department of Physics, National Sun Yat-Sen University, No. 70, Lien-Hai Road, Kaohsiung City 80424, Taiwan, R.O.C.}
\affiliation{Center of Astronomy and Gravitation, National Taiwan Normal University, Taipei 116, Taiwan}

\author[0000-0002-0433-9840]{Simon Casassus}
\affil{Departamento de Astronom\'ia, Universidad de Chile, Casilla 36-D, Santiago, Chile}
\affil{Data Observatory Foundation, to Data Observatory Foundation, Eliodoro Ya\'n\~ez 2990, Providencia, Santiago, Chile}
\affil{Millennium Nucleus on Young Exoplanets and Their Moons (YEMS), Santiago, Chile}

\author[0000-0001-9290-7846]{Ruobing Dong}
\affil{Department of Physics \& Astronomy, University of Victoria, Victoria, BC, V8P 5C2, Canada}

\author[0000-0003-1958-6673]{Kiyoaki Doi}
\affiliation{Department of Astronomical Science, School of Physical Sciences, Graduate University for Advanced Studies (SOKENDAI), 2-21-1 Osawa, Mitaka, Tokyo 181-8588, Japan}
\affiliation{National Astronomical Observatory of Japan, 2-21-1 Osawa, Mitaka, Tokyo 181-8588, Japan}

\author[0000-0002-3053-3575]{Jun Hashimoto}
\affil{Astrobiology Center, National Institutes of Natural Sciences, 2-21-1 Osawa, Mitaka, Tokyo 181-8588, Japan}
\affil{Subaru Telescope, National Astronomical Observatory of Japan, Mitaka, Tokyo 181-8588, Japan}
\affil{Department of Astronomical Science, School of Physical Sciences, Graduate University for Advanced Studies (SOKENDAI), 2-21-1 Osawa, Mitaka, Tokyo 181-8588, Japan}


\author{Takayuki Muto}
\affil{Division of Liberal Arts, Kogakuin University, 1-24-2, Nishi-Shinjuku, Shinjuku-ku, Tokyo 163-8677, Japan}



\begin{abstract}
PDS~70 is a protoplanetary system that hosts two actively accreting gas giants, namely PDS~70b and PDS~70c. 
The system has a $\sim$60--100 au dusty ring that has been resolved by the Atacama Large Millimeter/Submillimeter Array (ALMA), along with circumplanetary disks around the two gas giants. 
Here we report the first Karl G. Jansky Very Large Array (JVLA) Q (40--48 GHz), Ka (29--37 GHz), K (18--26 GHz), and X (8--12 GHz) bands continuum observations, and the complementary ALMA Bands 3 ($\sim$98 GHz) and 4 ($\sim$145 GHz) observations towards PDS~70.
The dusty ring appears azimuthally asymmetric in our ALMA images.
We obtained firm detections at Ka and K bands without spatially resolving the source; we obtained a marginal detection at Q band, and no detection at X band.
The spectral indices ($\alpha$) are 5$\pm$1 at 33--44 GHz and 0.6$\pm$0.2 at 22--33 GHz.
At 10--22 GHz, the conservative lower limit of $\alpha$ is 1.7.
The 33--44 GHz flux density is likely dominated by the optically thin thermal emission of grown dust with $\gtrsim$1 mm maximum grain sizes, which may be associated with the azimuthally asymmetric substructure induced by planet-disk interaction.
Since PDS~70 was not detected at X band, we found it hard to explain the low spectral index at 22--33 GHz only with free-free emission. 
Hence, we attribute the dominant emission at 22--33 GHz to the emission of spinning nanometer-sized dust particles, while free-free emission may partly contribute to emission at this frequency range.
In some protoplanetary disks, the emission of spinning nanometer-sized dust particles may resemble the 20--50 GHz excess in the spectra of millimeter-sized dust.
The finding of strong continuum emission of spinning nanometer-sized particles can complicate the procedure of constraining the properties of grown dust.
Future high resolution, multi-frequency JVLA/ngVLA and SKA observations may shed light on this issue. 
\end{abstract}



\section{Introduction} \label{sec:intro}

\begin{deluxetable*}{l l l l r l r r}
\tablecaption{Observations\label{tab:jvla}}
\tablewidth{700pt}
\tabletypesize{\scriptsize}
\tablehead{
\colhead{Date} &
\colhead{Band} &
\colhead{API rms\tablenotemark{a}} &  
\colhead{{\it uv}-range} &
\colhead{Freq. range} &
\colhead{Flux/Passband/Gain calibrators} &
\colhead{Synthesized beam\tablenotemark{b}} &
\colhead{rms noise\tablenotemark{b}}  \\
\colhead{(UTC)} &
\colhead{} &
\colhead{($^{\circ}$)} & 
\colhead{(meters)} &
\colhead{(GHz)} &
\colhead{} &
\colhead{(\beam; $^{\circ}$)} &
\colhead{(\ujyperbeam)} 
} 
\startdata
\multicolumn{8}{c}{Project: JVLA (19B-023; PI: H. B. Liu)} \\
2019-Nov-08 &
X&
4--7 &
36-19070&
8--12 & 
3C286/3C286/J1427-4206&
16$\times$4\farcs9; $-$0.61$^{\circ}$&
10\\
2019-Nov-15 &
K&
4--6 &
33-1031&
18--26 & 
3C286/J1427-4206/J1427-4206&
32$\times$2\farcs4; $-$0.43$^{\circ}$&
17\\
2019-Nov-17 &
K&
$\sim$7 &
32-1030&
18--26 & 
3C286/J1427-4206/J1427-4206&
35$\times$2\farcs7; $-$0.19$^{\circ}$&
16\\
2020-Jan-04 &
Q&
$\sim$3 &
34-1031&
40--48 & 
3C286/3C286/J1427-4206&
9\farcs6$\times$1\farcs2; $-$0.35$^{\circ}$&
100\\
2020-Jan-20 &
Ka&
$\sim$4 &
36--3381&
29--37 & 
3C286/3C286/J1427-4206&
7\farcs0$\times$3\farcs0; $-$14$^{\circ}$&
23\\
\hline
\multicolumn{8}{c}{Project: ALMA (2022.1.01477.S; PI: H. B. Liu)} \\
2023-Mar-26 &
3 &
$\cdots$ &
14--1241&
91.5--95.2; 101.6--105.3 & 
J1427-4206/J1427-4206/J1407-4302&
0\farcs49$\times$0\farcs45; $-$30$^{\circ}$&
18\\
2023-Mar-25 &
4 &
$\cdots$ &
14--1246&
140.0--142.7; 149.0--152.8 & 
J1427-4206/J1427-4206/J1407-4302&
0\farcs43$\times$0\farcs33; 44$^{\circ}$&
44\\
\enddata
 \tablenotetext{a}{The RMS phase at the JVLA site measured with the Atmospheric Phase Interferometer (API; for more details see \url{https://science.nrao.edu/facilities/vla/docs/manuals/oss2013A/performance/gaincal/api}). The API a 2-element interferometer separated by 300 meters, observing an 11.7 GHz beacon from a geostationary satellite. The default API rms upper limits for the X, Ku, K, Ka, and Q band observations are 30$^{\circ}$, 15$^{\circ}$, 10$^{\circ}$, 7$^{\circ}$, and 5$^{\circ}$.}\vspace{-0.2cm}
 \tablenotetext{b}{Measured from multi-frequency synthesis images generated using aggregated continuum bandwidths and Natural (i.e., Briggs Robust$=$2) weighting.}
\end{deluxetable*}

Planet-disk interaction plays an important role in trapping grown dust, which could potentially aid in dust growth thereby facilitate the formation of second-generation planetesimals and planets (for a recent review see \citealt{Birnstiel2023arXiv231213287B} and references therein).
Constraining dust distributions and dust growth in planet-hosting protoplanetary disks is crucial for understanding the related physical processes.
Limited by the sensitivity and image fidelity of the present optical/infrared observing facilities, the candidates of planet-hosting protoplanetary disks remain rare; most of those candidates still require confirmation (e.g., \citealt{Kraus2012ApJ...745....5K,Currie2022NatAs...6..751C}). 

The pre-main sequence star PDS~70 is a K7-type (0.76 $M_{\odot}$), weak-line T Tauri type star (\citealt{Keppler2018A&A...617A..44K,Muller2018A&A...617L...2M}) at $\sim$113 pc distance (\citealt{Gaia_2023}).
The previous Subaru-HiCIAO coronagraphic polarimetric imaging observations at 1.6 $\mu$m wavelength resolved that it is associated with an extended disk of $\sim$100 au radius, which presents a gap extending from $\sim$17--60 au radii (0$''$.15--0$''$.53; \citealt{Hashimoto2012ApJ...758L..19H}).
In addition, \citet{Hashimoto2015ApJ...799...43H} compared the Subaru-HiCIAO near infrared images with the 1.3 mm image taken with the Submillimeter Array (SMA).
Based  detailed Monte Carlo radiative transfer models (also see \citealt{Dong2012ApJ...760..111D}), they suggested that there are at least two distinct dust components in the disk which have large and small grain sizes.
It appeared that the gap edges of the large and small dust components are at different radii, which are 80 au and 65 au, respectively.
This may be a signpost of dust filtration, which may be due to the presence of (proto)planets.

Newer optical and infrared imaging observations equipped with extreme adaptive optics have discovered the 2--17 Jupiter masses ($M_{\mbox{\scriptsize Jup}}$) planet orbiting PDS~70, known as PDS~70b (\citealt{Keppler2018A&A...617A..44K,Muller2018A&A...617L...2M,Hashimoto2020AJ....159..222H,Wagner2018ApJ...863L...8W}), followed by the detection of another 4--12 $M_{\mbox{\scriptsize Jup}}$ planet, PDS~70c (\citealt{Haffert2019NatAs...3..749H,Hashimoto2020AJ....159..222H}).
These two planets orbit the host pre-main sequence star at 23 au (0\farcs2) and 35 au (0\farcs31) radii, respectively.
The resolved massive planets at known locations make PDS70 a unique and ideal laboratory for investigating dust processing in a planet-hosting disk using (sub)millimeter and centimeter wavelength observations.


Intensive Atacama Large Millimeter/Submillimeter Array (ALMA) dust continuum observations at 225--345 GHz frequencies ($\lambda\sim$1.3--0.87 mm) have been carried out towards PDS~70 (\citealt{Long2018ApJ...858..112L,Isella2019ApJ...879L..25I,Benisty2021ApJ...916L...2B,Facchini2021AJ....162...99F,Casassus2022MNRAS.513.5790C}).
They resolved the gap/ring structures in the PDS~70 disk. 
Moreover, they detected the circumplanetary disks around PDS~70b and c (\citealt{Isella2019ApJ...879L..25I,Benisty2021ApJ...916L...2B,Casassus2022MNRAS.513.5790C}). 
These observations have provided the unprecedentedly clear picture of the structures of the bulk of the dust reservoirs in the PDS~70 system. 
However, the recent surveys and population synthesis studies have indicated that the 230--345 GHz optical depths of most of the Class II protoplantary disks in the Taurus, Ophiuchus, and Lupus star-forming regions are $\gtrsim$5--10 (\citealt{Chung2024arXiv240519867C,Delussu2024arXiv240514501D}, and references therein). 
In addition, some case studies have shown that a good amount of grown dust may be hidden in localized substructures that can only be discerned at $\lesssim$100 GHz frequencies (e.g., \citealt{Hashimoto2022ApJ...941...66H,Hashimoto2023AJ....166..186H,Liu2024A&A...685A..18L}).
Continuum observations at yet lower frequencies, reaching into the microwaves,  will reach an optically thinner regime, and will provide constraints on the dust mass and on the properties of the largest grains observables. Such observations of grain growth are crucial to understand the  planet-formation process. 

\begin{figure*}
    \hspace{-2cm}
    \begin{tabular}{ p{4.3cm} p{4.3cm} p{4.3cm} p{4.3cm}}
    \includegraphics[height=5.3cm]{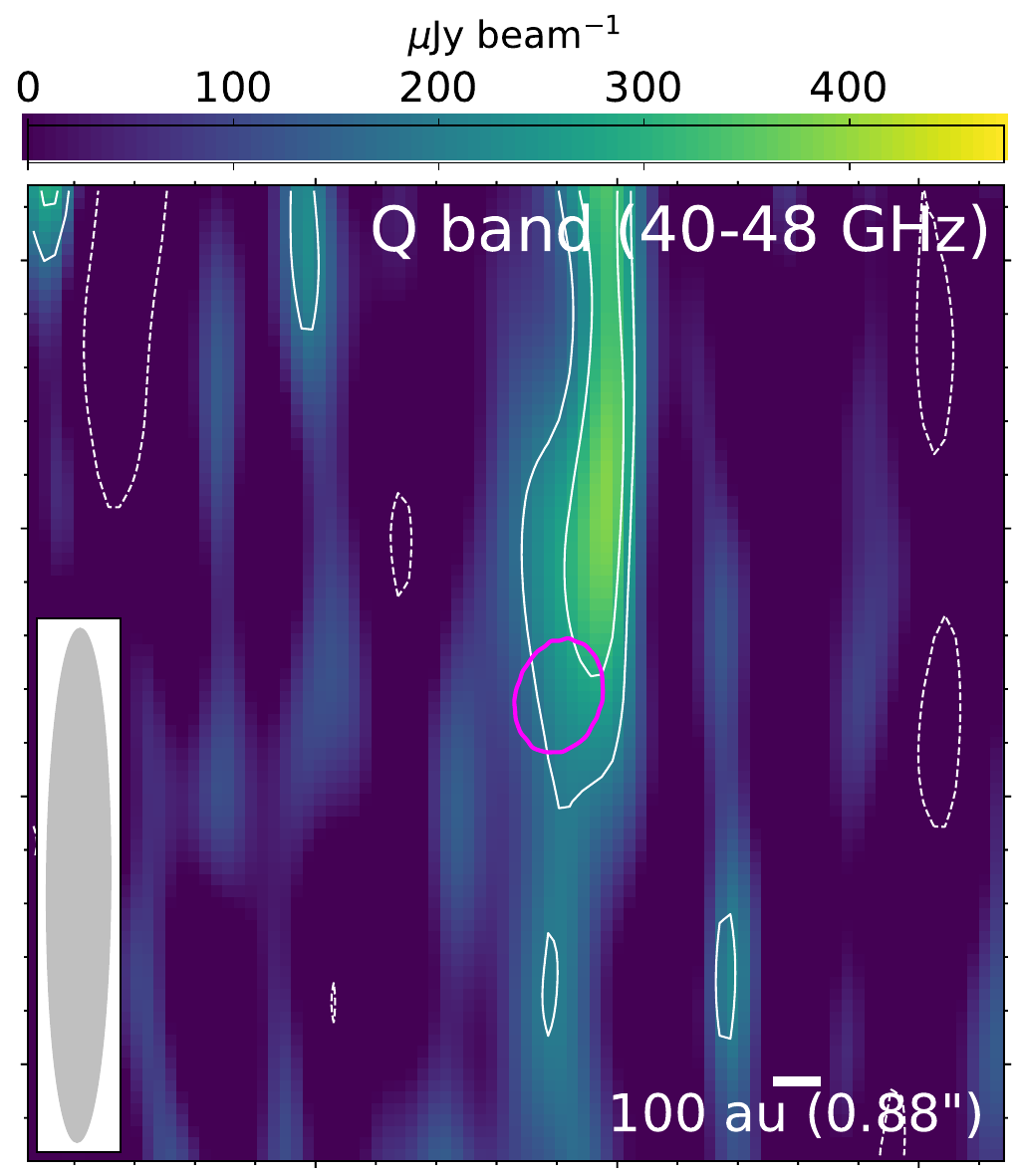} &
    \includegraphics[height=5.3cm]{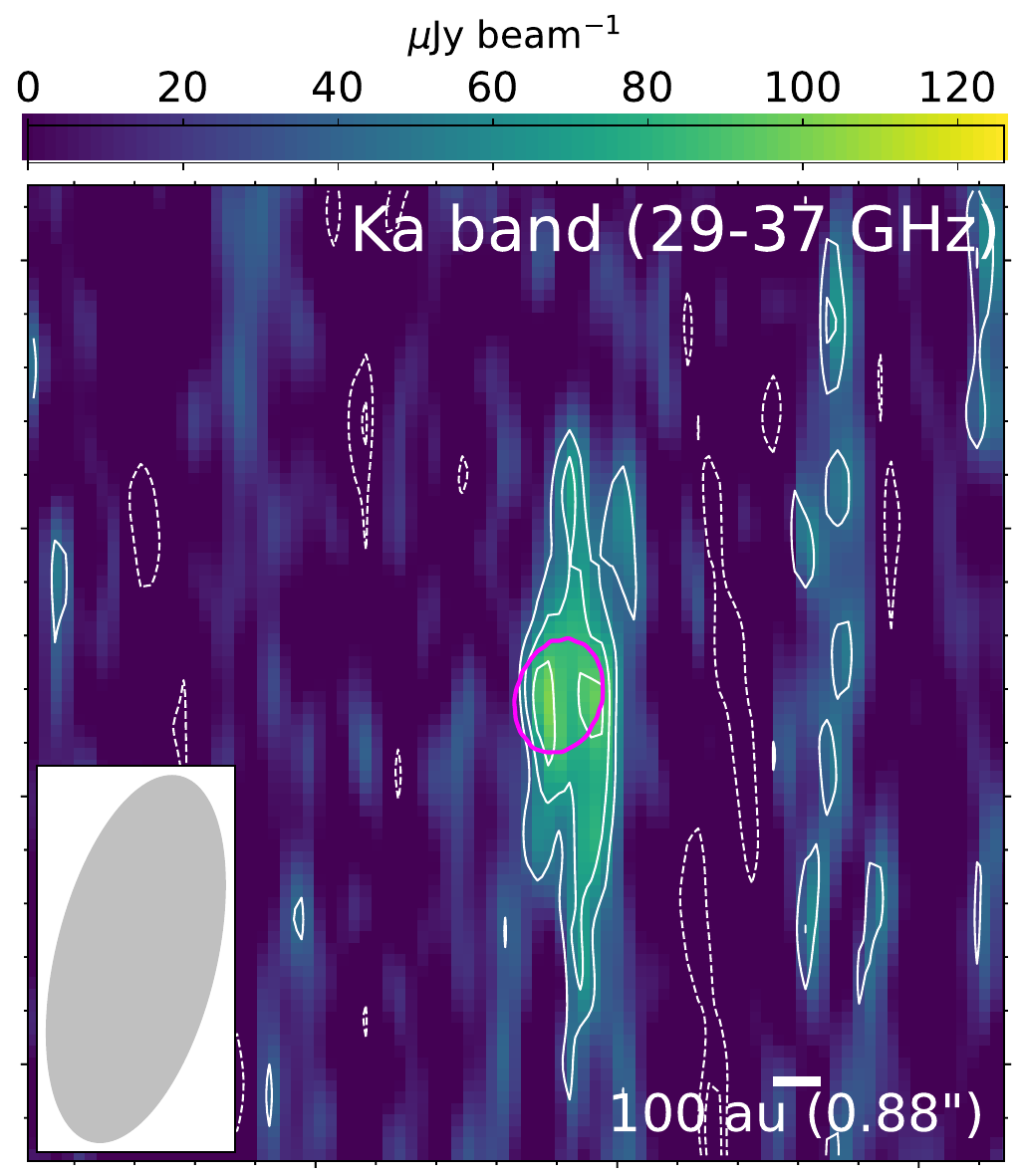} &
    \includegraphics[height=5.3cm]{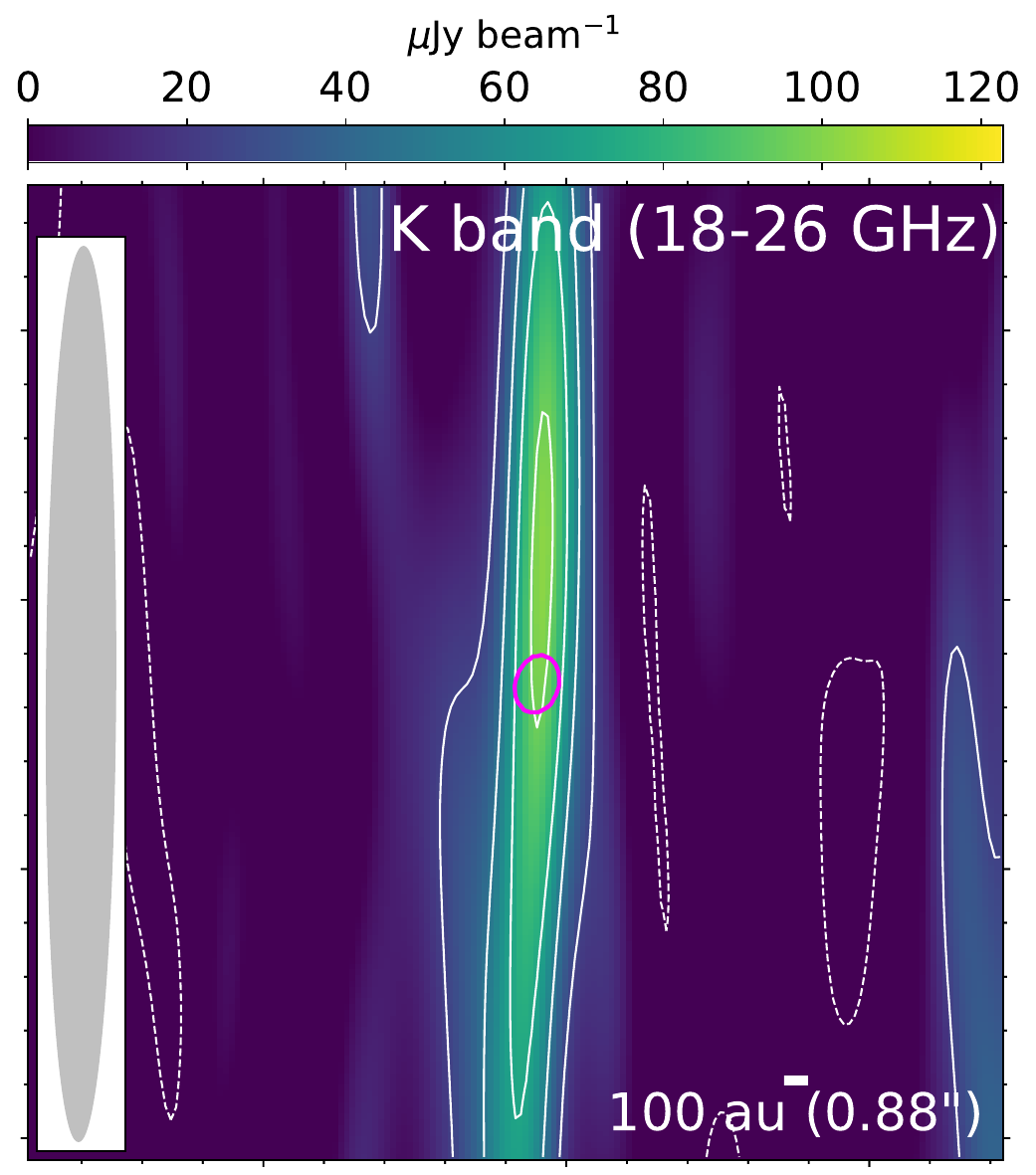} &
    \includegraphics[height=5.3cm]{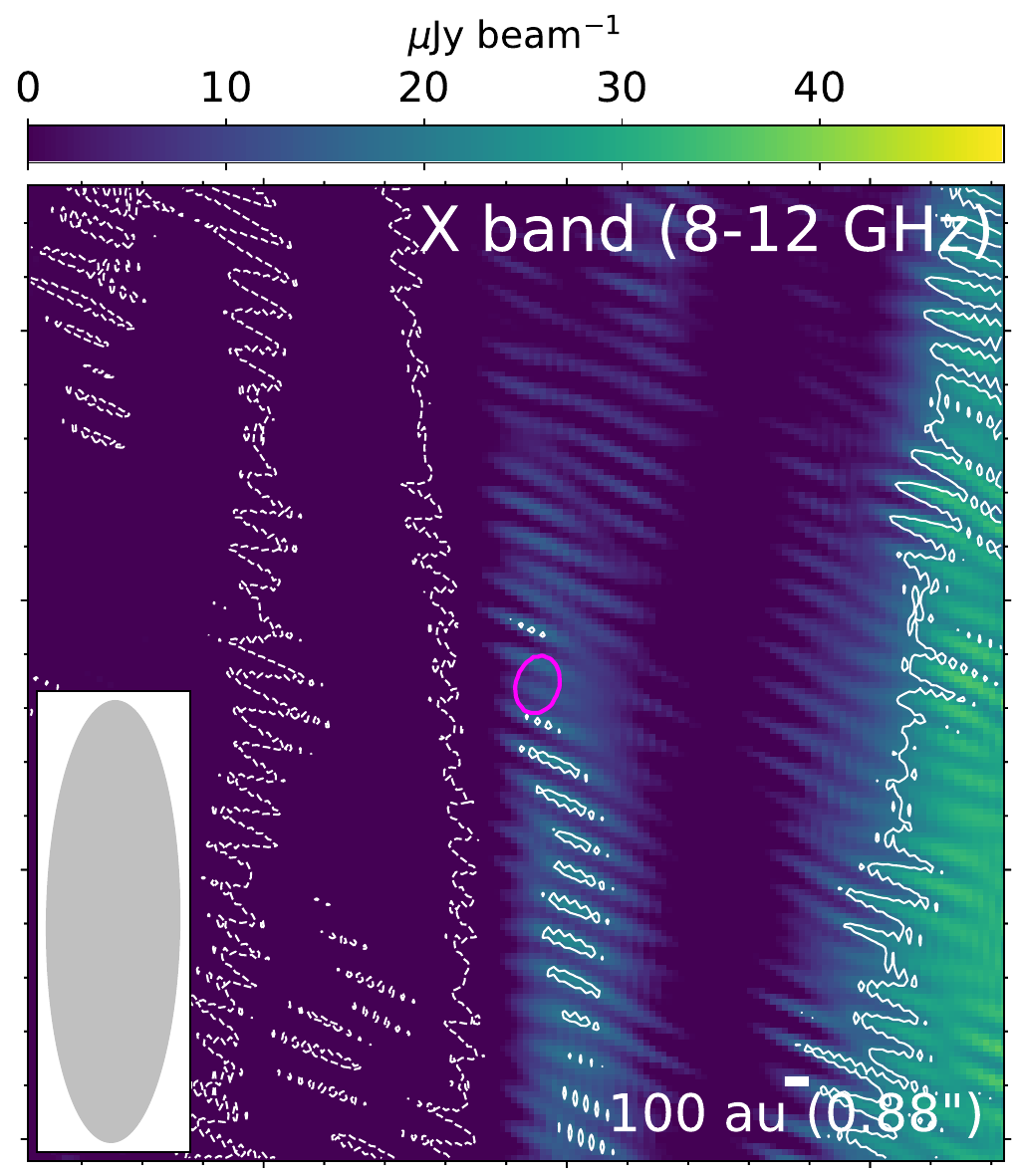} \\
    \includegraphics[height=5.3cm]{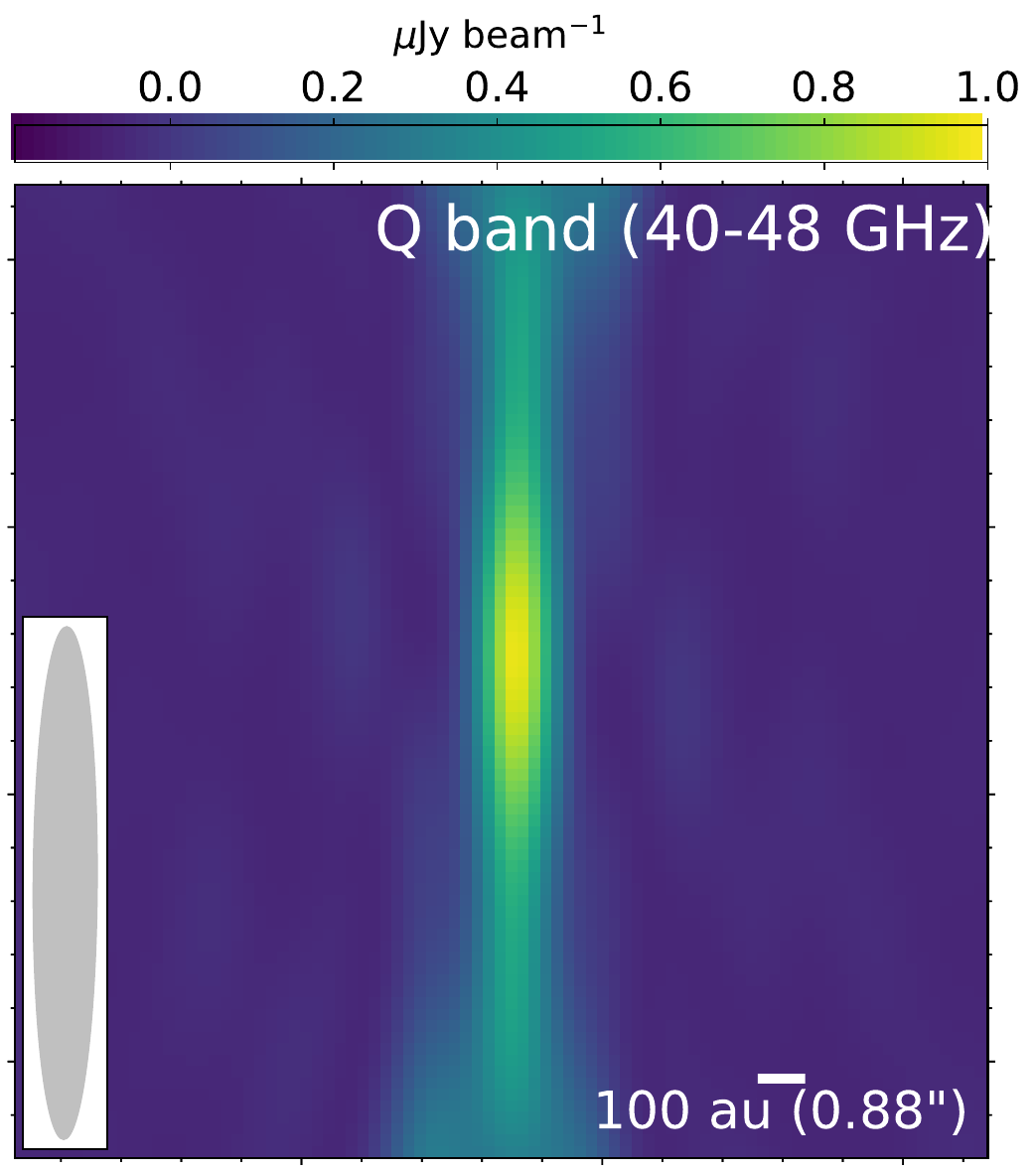} &
    \includegraphics[height=5.3cm]{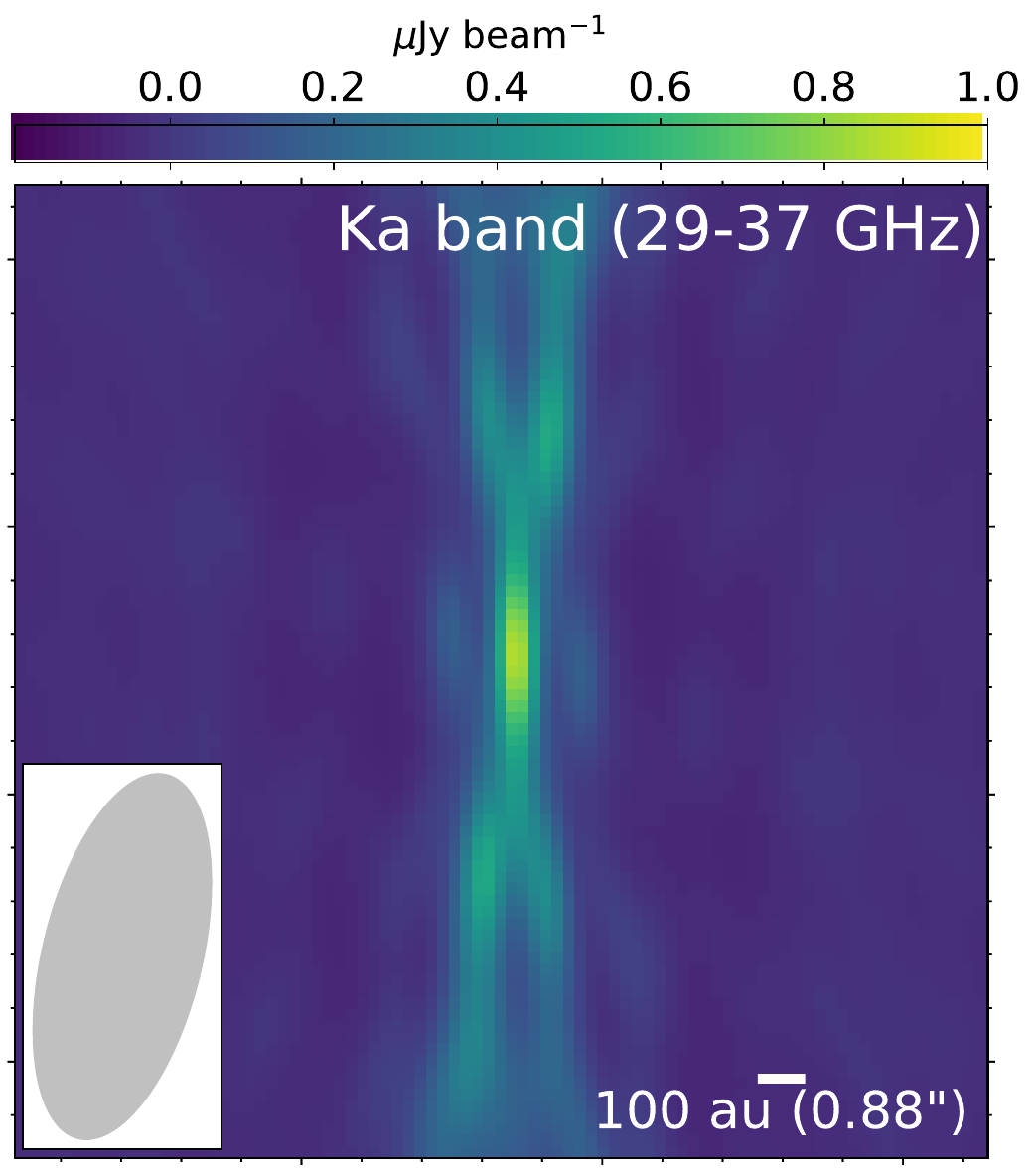} &
    \includegraphics[height=5.3cm]{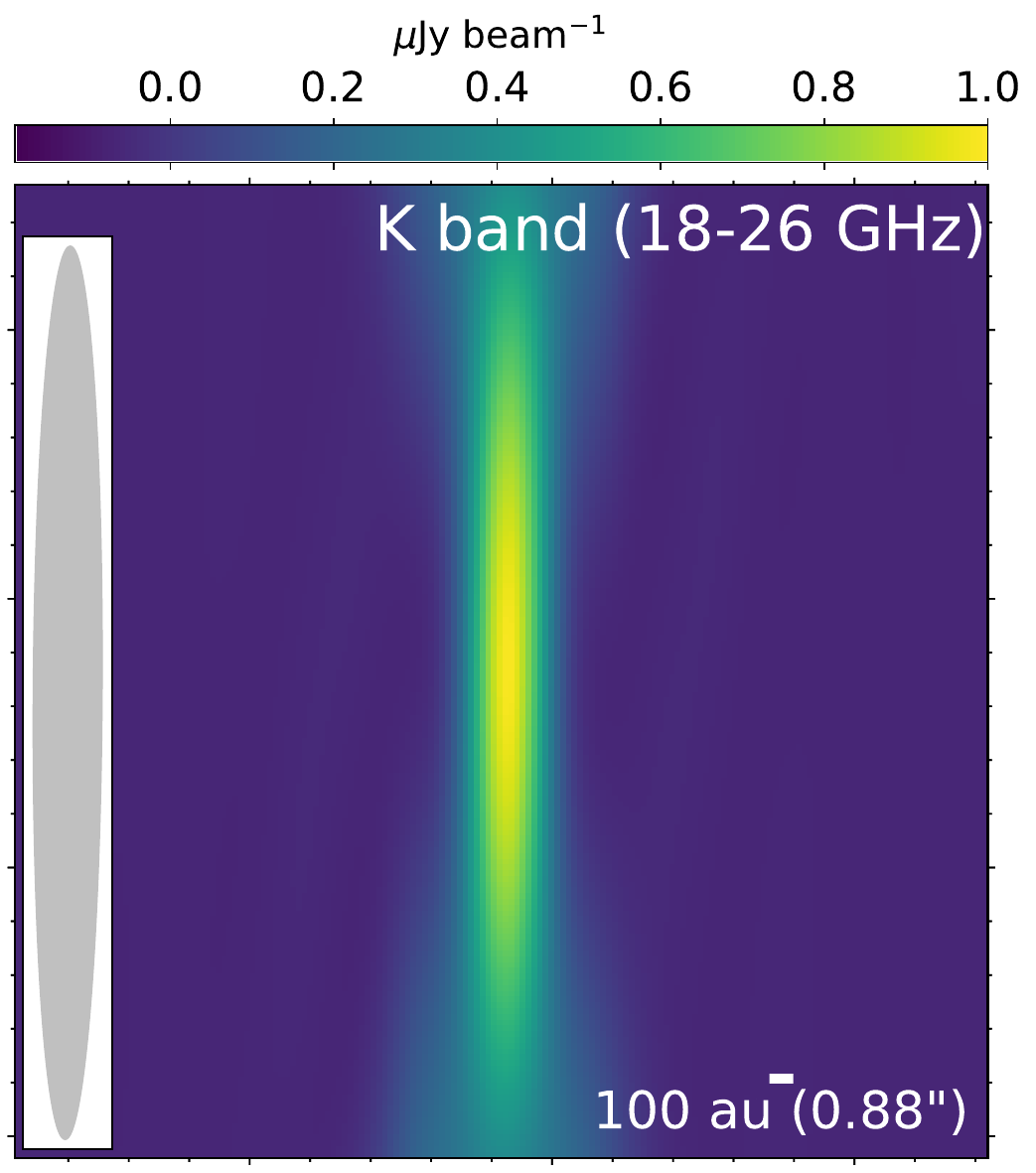} &
    \includegraphics[height=5.3cm]{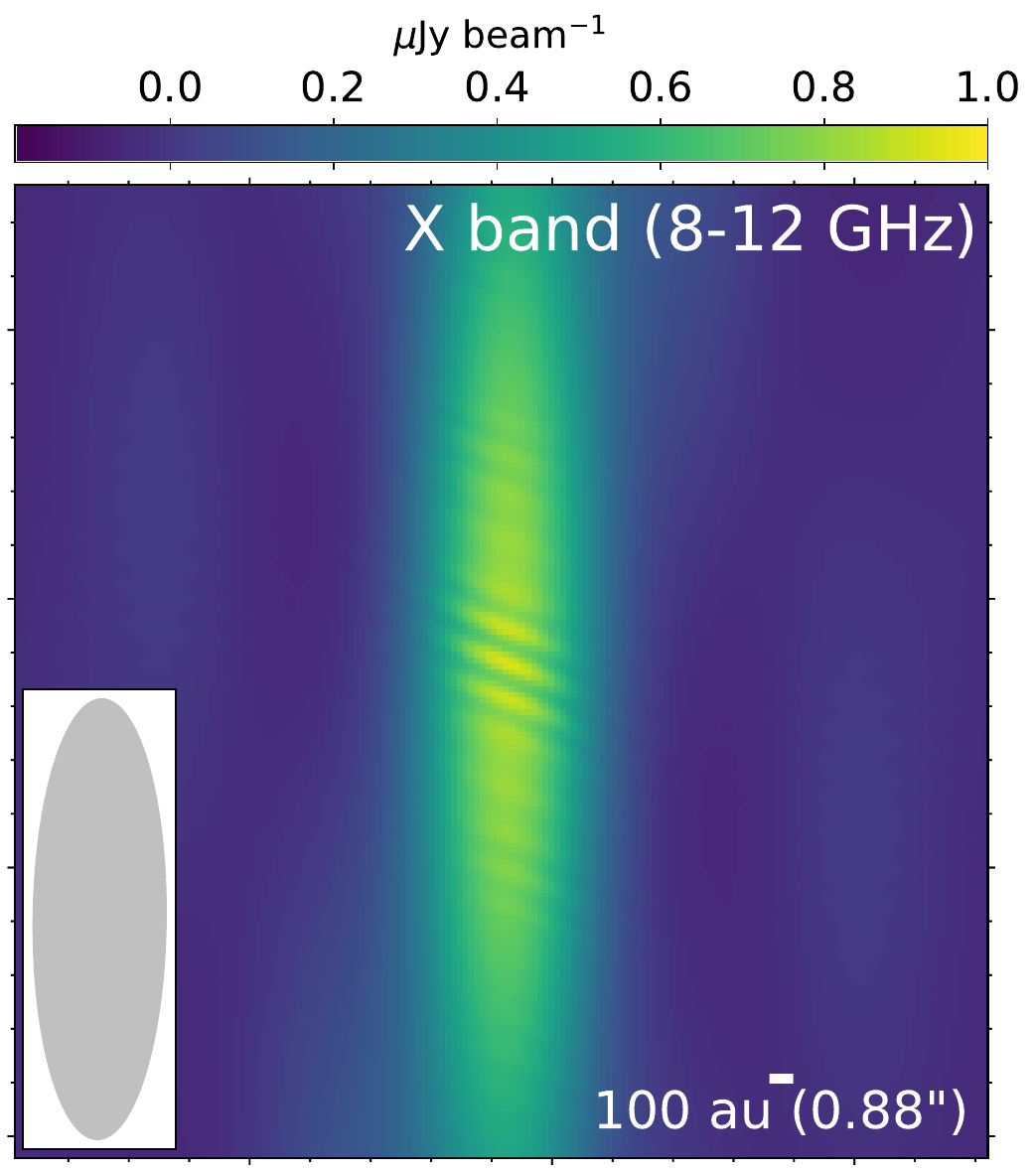} \\
    \end{tabular}
    \caption{
    JVLA continuum images (upper row, color images and contours) and the synthesized dirty beams (bottom row, color images).
    {\it Upper row :--} From left to right are the Q band (40--48 GHz) image (contours are (1-$\sigma$) $\times$[-2, 2, 3]); Ka band (29--37 GHz) image (contours are (1-$\sigma$) $\times$[-3, -2, 2, 3, 4]); K band (18--26 GHz) image (contours are (1-$\sigma$) $\times$[-2, 2, 4, 6, 8]) that were generated by jointly imaging the two epochs of K band observations (Table \ref{tab:jvla}); X band (8--12 GHz) image (contours are (1-$\sigma$) $\times$[-2, 2]). The synthesized clean beams are shown in the lower left. 
    We note that the two panels on the left and the two panels on the right are presented on different angular scales due to the very elongated synthesized (clean) beams of the latter cases. 
    The magenta ellipse in each panel shows the location and size of the PDS~70 millimeter ring (c.f. \citealt{Long2018ApJ...858..112L,Isella2019ApJ...879L..25I,Benisty2021ApJ...916L...2B,Facchini2021AJ....162...99F,Casassus2022MNRAS.513.5790C}).
    {\it Bottom row :--} Color images show the synthesized dirty beams of the observations presented in the upper row.
    }
    \vspace{0.37cm}
    \label{fig:image}
\end{figure*}

Here we report our JVLA (18--48 GHz) observations and the ALMA Bands 3 (97.5 GHz) and 4 (145 GHz) observations towards PDS~70.
The technical details of our observations are provided in Section \ref{sec:observations}.
The results are introduced in Section \ref{sec:results}.
Section \ref{sec:discussion} discusses our hypotheses of the dominant emission mechanisms at 18--48 GHz.
Our conclusion is given in Section \ref{sec:conclusion}.
Appendix \ref{appendix:flux} briefly discusses various strategies of measuring flux densities. 
Appendix \ref{appendix:model} described how we produced models of dust spectral energy distributions (SEDs) to compare with the JVLA observations.

\section{Observations and data reduction}\label{sec:observations}
\subsection{JVLA}\label{sub:JVLA}

We have carried out the JVLA observations towards PDS~70 at Q (40--48 GHz), Ka (29--37 GHz), K (18--26 GHz), and X (8--12 GHz) bands in 2019 (Project code: 19B-023, PI: Hauyu Baobab Liu), which are summarized in Table~\ref{tab:jvla}.
The pointing and phase referencing centers were R.A. (J2000)$=$14\ras08\ram10\farcs150, Decl. (J2000)$=-$41\decd23\decm52\farcs5. 
We adopted the standard 3-bit continuum setup of the WIDAR correlator, which provided full RR, RL, LR, and LL correlations.

The K and Q bands observations were carried out in the most compact (D) array configuration.
The Ka band observations were carried out in the D-north-C (DnC) array configuration that the western and eastern arms of the array were in the D configuration while the northern arm of the array was in the C array configuration.
The X band observations were carried out in a moving array configuration that most of the antennae were on the pads of the D array configuration while three antennae were on the pad of the A array configuration.
The target source is in the far south which transited at a $\sim$14.6$^{\circ}$ elevation.
In all of the observations, the antennae in the northern arm of the array were severely shadowed and thus mostly had to be flagged.
Effectively, we were observing with an array with only the western and eastern arm, which poorly determined the location of the target source in the north-south direction.
After flagging the shadowed and malfunctioned antennae, in general, we were left with $\sim$15 available antennae.

\begin{figure}
    \hspace{-2cm}
    \begin{tabular}{ p{4.3cm} p{4.3cm} p{4.3cm} p{4.3cm}}
    \includegraphics[height=5.3cm]{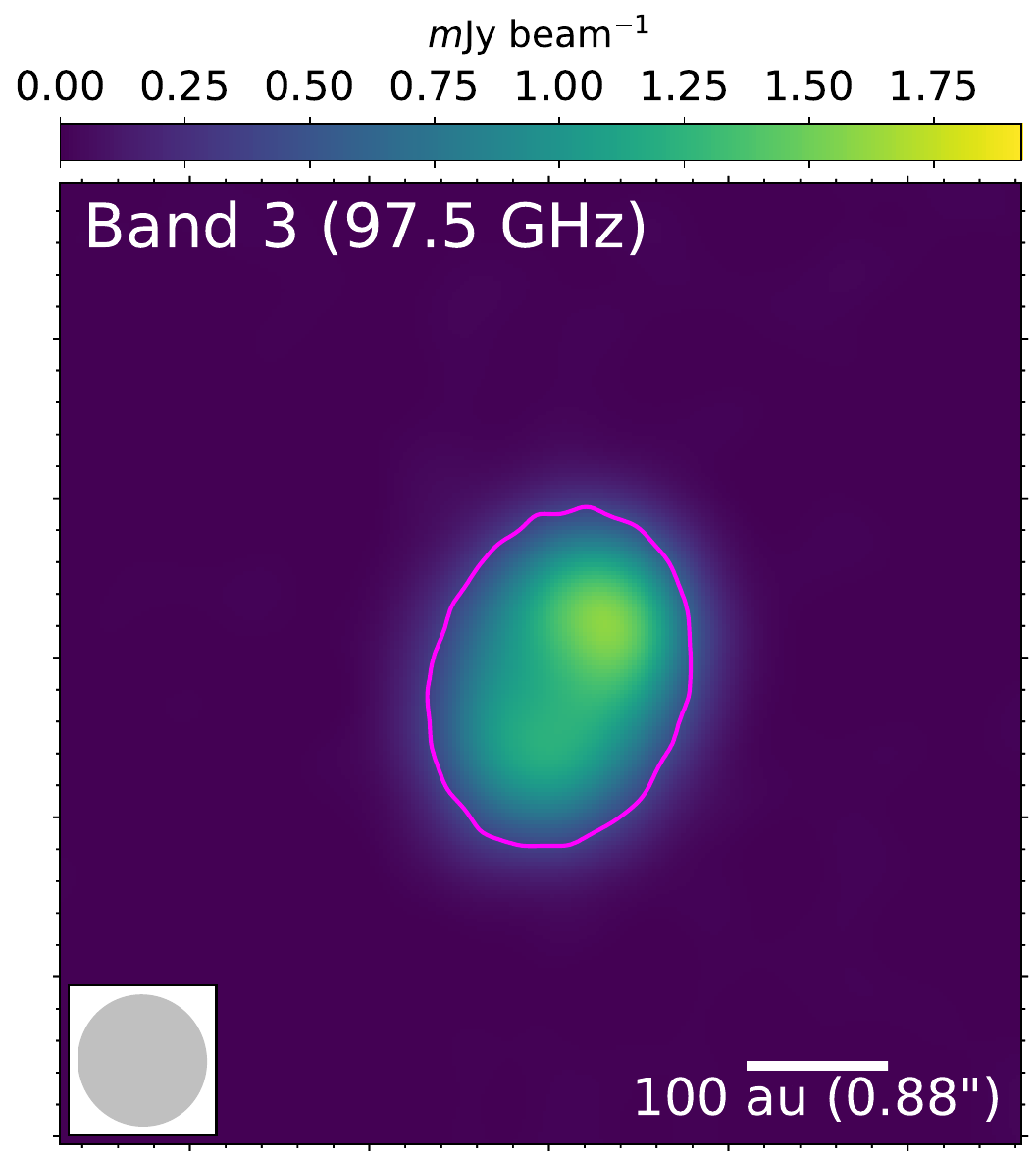} &
    \includegraphics[height=5.3cm]{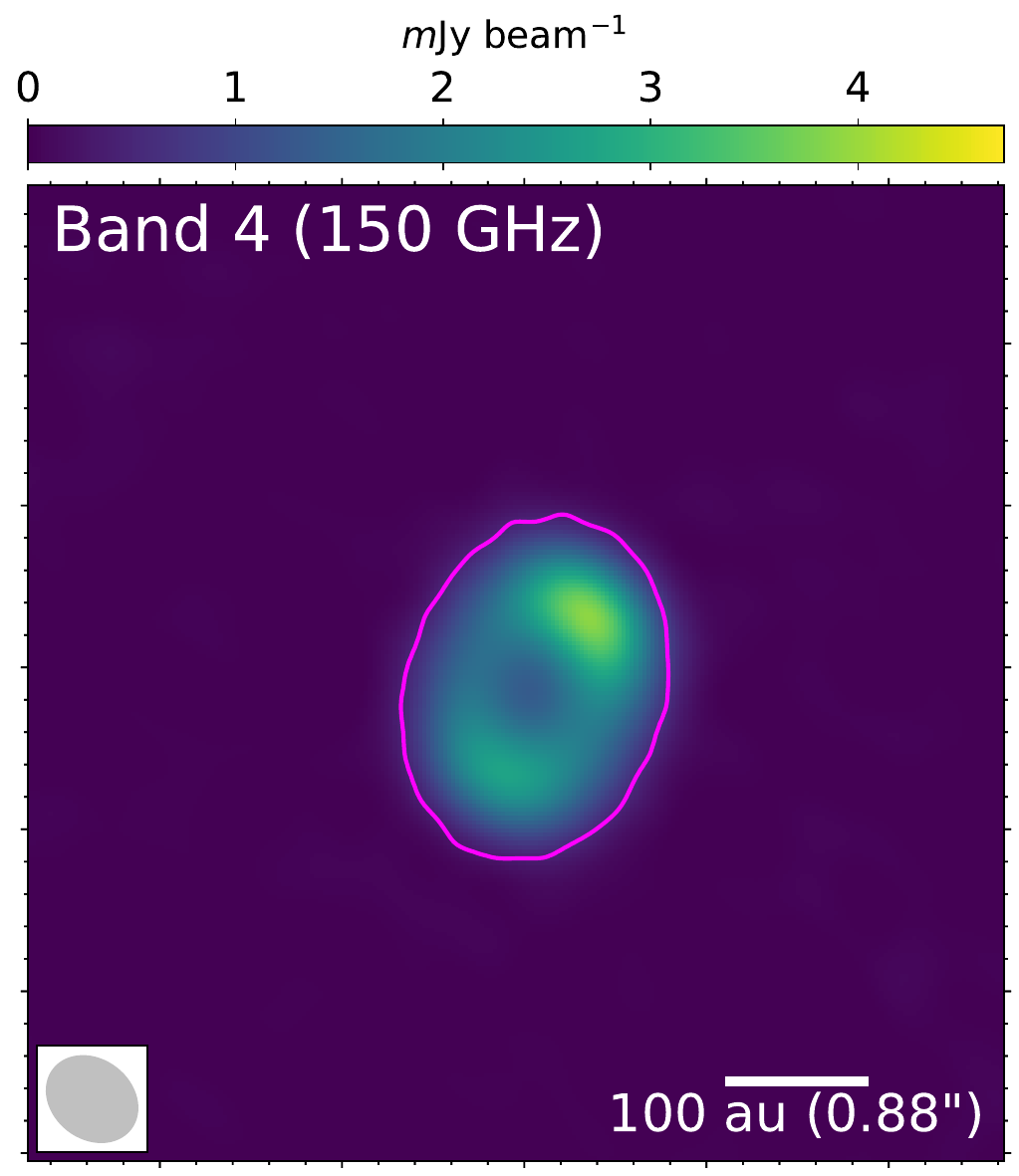} \\
    \end{tabular}
    \caption{
    ALMA continuum images at Bands 3 (left) and 4 (right).
    Synthesized beams of these images are shown in the lower left. 
    Magenta ellipse are the same as those in Figure \ref{fig:image}.
    }
    \label{fig:almaimage}
\end{figure}

We manually calibrated the JVLA data following the standard strategy (see below) using the Common Astronomy Software Applications (CASA 6.4.1; \citealt{McMullin2007ASPC..376..127M}) software package.
For all the JVLA observations, after implementing the antenna position corrections, weather information, gain-elevation curve, and opacity model, we bootstrapped delay fitting and passband calibrations, and then
performed complex gain calibration. 
We applied the absolute flux reference to our complex gain solutions, and then applied all derived solution tables to the target source.

We performed multi-frequency synthesis (mfs; nterms$=$1) imaging (\citealt{Rau2011A&A...532A..71R}) using the CASA \texttt{tclean} task.
We adopted the Briggs Robust$=$2.0 weighting to minimize the root-mean-square (RMS) noises in Jy\,beam$^{-1}$ units.
In the imaging procedure, the dirty images were first obtained from the inverse Fourier transform of the Robust$=$2.0 weighted visibilities; the synthesized dirty beams were the inverse Fourier transforms of the sampling functions in the visibility domain.
The \texttt{tclean} task used the synthesized dirty beams to deconvolve the dirty images and at the end outputed the deconvolution model (i.e., the \texttt{clean} components) and the \texttt{clean} residual image. 
The \texttt{tclean} task defined the synthesized \texttt{clean} beam by performing two-dimensional (2D) Gaussian fittings to the synthesized dirty beam.
Finally, the \texttt{tclean} task produced the \texttt{clean} images by convolving the deconvolution model with the synthesized \texttt{clean} beams, and then superimposing the \texttt{clean} residual image.
Table \ref{tab:jvla} summarizes the achieved synthesized \texttt{clean} beams and the root-mean-square (RMS) noises that were measured from the \texttt{clean} residual images made for the individual epochs of observations.
We jointly imaged the two epochs of K band observations, which yielded a \beam=32$''\times$2\farcs4 (P.A.=$-$0.43$^{\circ}$) synthesized beam and a RMS noise of 12 $\mu$Jy\,beam$^{-1}$.

The synthesized beam of the X band image appears grainy due to the three antennae on the A array configuration pads (Figure \ref{fig:image}). 
The synthesized beam of the Ka band image appeared highly non-Gaussian in the central region and thus cannot be well represented by the 2D Gaussian synthesized \texttt{clean} beams (Figure \ref{fig:image}). 
The {\tt clean} algorithm can still deconvolve the dirty image in this case, although the interpretation of the restored image and the Jy\,beam$^{-1}$ intensity unit are less trivial (Appendix \ref{appendix:flux}).

\subsection{ALMA}\label{sub:ALMA}
We have carried out the ALMA Bands 3 (89.6--93.4 GHz/101.6--105.4 GHz) and 4 (137.0--138.9 GHz/149.0--152.9 GHz) standard FDM continuum observations in the C43-5 array configuration (Project code: 2022.1.01477.S, PI: Hauyu Baobab Liu; Table~\ref{tab:jvla}).
The pointing and phase referencing centers were R.A. (ICRS)$=$14\ras08\ram10\farcs1067, Decl. (ICRS)$=-$41\decd23\decm53\farcs059. 

We manually calibrated the ALMA data following the standard strategy that is similar to that for the JVLA data calibration (Section \ref{sub:JVLA}) using CASA 6.4.1 (\citealt{McMullin2007ASPC..376..127M}).
In addition, we performed gain-phase self-calibration. 
We produced continuum data by binning the line-free channels.
We performed mfs (nterms$=$1) imaging (\citealt{Rau2011A&A...532A..71R}) for the continuum data using the CASA \texttt{tclean} task.
We adopted the Briggs Robust$=$2.0 weighting to minimize the RMS noises in Jy\,beam$^{-1}$ units.
The achieved image qualities are summarized in  Table~\ref{tab:jvla}.

\begin{figure}
    \vspace{-0.3cm}
    \hspace{-1.2cm}
    \begin{tabular}{p{9cm}}
    \includegraphics[width=9.1cm]{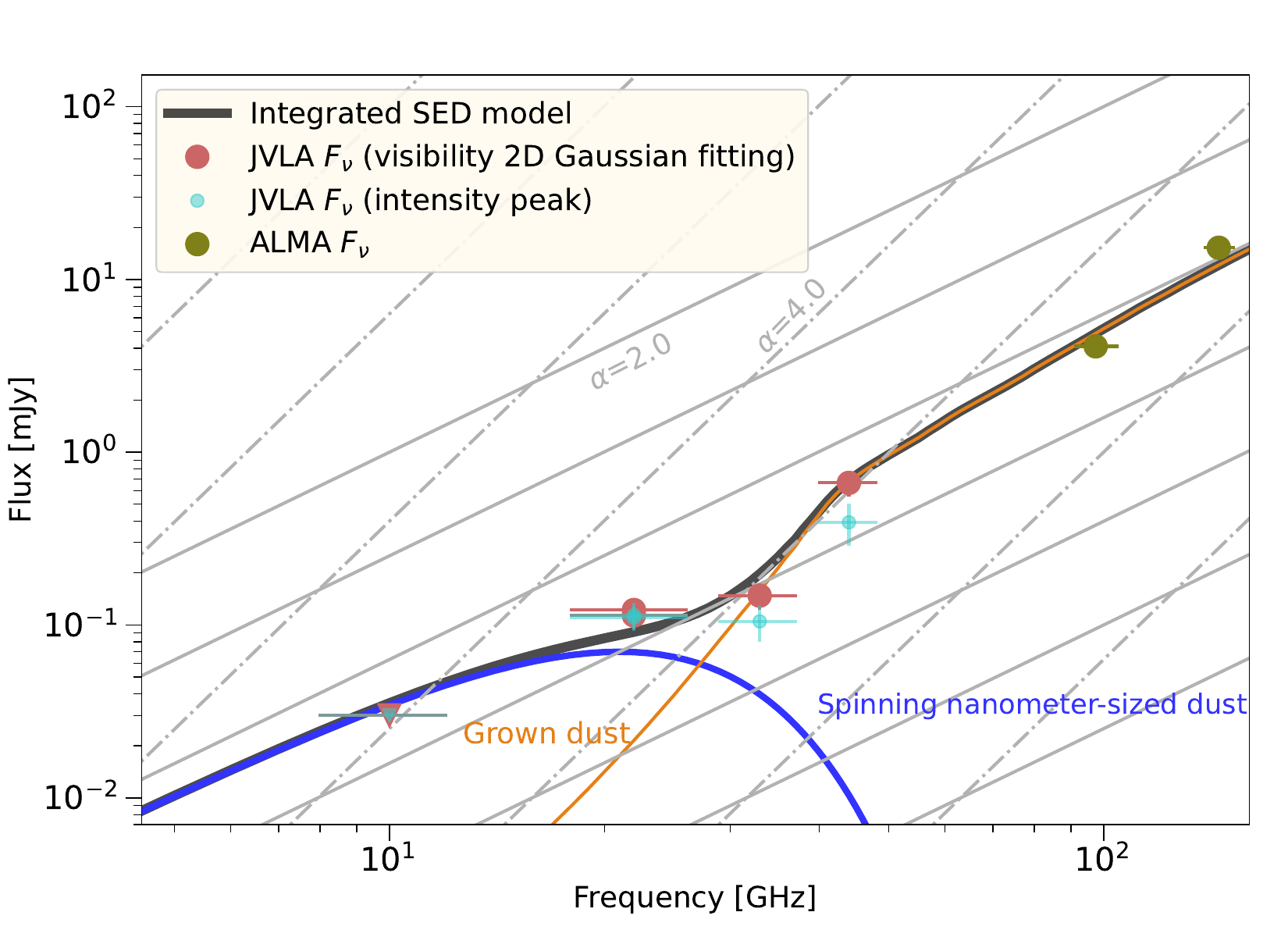} \\
    \vspace{-1.0cm}\includegraphics[width=9.1cm]{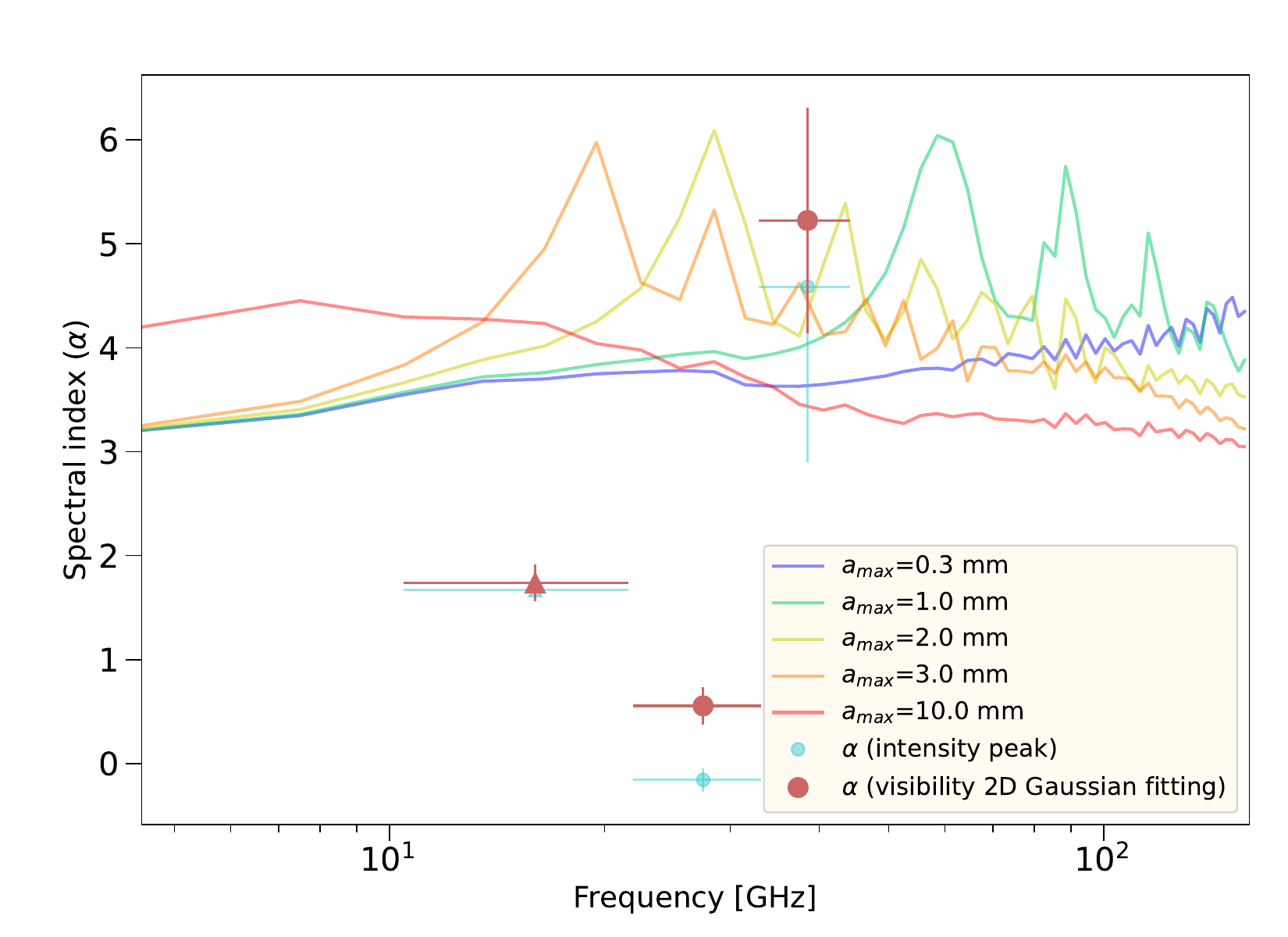} \\
    \end{tabular}
    \caption{
    The 8--48 GHz spectral profile of PDS~70.
    {\it Upper panel :--} Flux densities measured based on fitting 2D Gaussians in the visibility domain (red dots) and reading peak intensities in the image domain (cyan dots); triangles show the 3-$\sigma$ upper limits at 10 GHz. The sizes of some symbols are larger than the error bars. Gray solid and dash-dotted lines show the spectral indices of 2.0 and 4.0, respectively. Orange and blue lines show the SED models for a grown ($a_{\mbox{\scriptsize max}}=$1.5 mm) dust component and a spinning nanometer-sized dust component, respectively (Appendix \ref{appendix:model}); black line shows the integrated SED of these two dust components. 
    {\it Bottom panel :--}
    The 22--33 GHz and 33--44 GHz spectral indices and the lower limit of the 10--22 GHz spectral index derived based on the measurements presented in the upper panel. They are compared with the spectral indices of optically thin dust slabs at 20 K temperature with various $a_{\mbox{\scriptsize max}}$ values, which were evaluated based on the DSHARP-default dust opacities (\citealt{Birnstiel2018ApJ...869L..45B}).
    }
    \label{fig:flux}
\end{figure}

\section{Results} \label{sec:results}

Upper row of Figure \ref{fig:image} and Figure \ref{fig:almaimage} show the \texttt{clean} images obtained from our JVLA and ALMA observations, respectively. 
The dirty beams of the JVLA observations are shown in the bottom row of  Figure \ref{fig:image}. 

The ALMA Band 3 and 4 observations marginally resolved the PDS~70 ring (Figure \ref{fig:almaimage}).
Intriguingly, the PDS~70 ring appears azimuthal asymmetric in these ALMA images, showing a bright crescent in the northwest (c.f. \citealt{Benisty2021ApJ...916L...2B}).

We detected PDS 70 at Ka (33 GHz) and K bands (22 GHz) with a significance of at least 4-$\sigma$ and 8-$\sigma$, respectively.
At Q band (44 GHz), the $\sim$4-$\sigma$ intensity peak is 3\farcs2 offset from the PDS~70 ring, which is smaller than the $\sim$10$''$ FWHM of a \texttt{clean} beam.
The offset of the Q band intensity peak is attributed to the residual phase and delay errors that are inevitable in high-frequency observations at low elevations. 
We did not detect PDS~70 at $\gtrsim$2-$\sigma$ significance at X band (10 GHz).

We measured the flux densities from the ALMA images by integrating the pixel values over the inner 4$''$ square region.
Thanks to the high signal-to-noise ratios achieved by the ALMA observations, varying the location and shape of the square region does not change the obtained flux density by more than the nominal 5\% absolute flux errors we quoted. 

We compared two methods for measuring flux densities from the JVLA images: (i) reading the peak intensities in the image domain, and (ii) performing 2D Gaussian fittings in the visibility domain.
We considered the peak intensities as lower limits of flux densities because they are less immune to imaging artifacts. 
We considered the flux densities obtained from the 2D Gaussian fittings as upper limits because they are potentially confused by unidentified background/foreground sources. 
We provide further discussion about flux density measurements in Appendix \ref{appendix:flux}.

The flux density measurements are summarized in Table \ref{tab:flux}.
As we are not aware of any strong confusion source within the field of view of our JVLA observations, we think that the flux densities measured using visibility fittings are closer to the actual flux densities. 
We have inspected the 4.85 GHz and 0.834 GHz images taken with the Parkes-MIT-NRAO (PMN) radio continuum surveys (\citealt{Griffith1993AJ....105.1666G}) and the Sydney University Molonglo Sky Survey (SUMSS; \citealt{Mauch2003MNRAS.342.1117M}) and did not find strongly confusing radio sources, although the rms noises of these surveys were relatively high ($\sim$1 mJy\,beam$^{-1}$).
Moreover, the adjacent infrared sources detected by the Wide-field Infrared Survey Explorer (WISE)\footnote{Obtained from the NASA/IPAC Extragalactic Database (NED). The NASA/IPAC Extragalactic Database (NED) is funded by the National Aeronautics and Space Administration and operated by the California Institute of Technology.} are $\gtrsim$1$'$ away, whose radio emission unlikely can confuse our JVLA Q and Ka bands observations. 
Our discussion (Section \ref{sec:discussion}) will mainly focus on measurements made by visibility fittings, although the methods of measuring flux density do not lead to qualitative differences in our discussion and conclusion.

Upper panel of Figure \ref{fig:flux} summarizes the flux density measurements, while the bottom panel shows the spectral indices derived from these measurements. 
The spectral indices ($\alpha$) are 5$\pm$1 and 0.6$\pm$0.2 at 33--44 GHz and 22--33 GHz, respectively.
We were surprised by the bright 22 GHz emission, the low 22--33 GHz spectral index, and the 10 GHz non-detection (Figure \ref{fig:flux}).
Based on the 22 GHz flux density and the 3-$\sigma$ upper limit of the 10 GHz flux density (Table \ref{tab:jvla}), we derived a lower limit of the 10--22 GHz spectral index to be 1.7.
Our working hypothesis to interpret this SED is provided in Section \ref{sub:interpretation}.


\section{Discussion} \label{sec:discussion}

\subsection{Interpreting the SED}\label{sub:interpretation}

\begin{deluxetable}{l c c}
\tablecaption{Flux density measurements\label{tab:flux}}
\tablewidth{700pt}
\tabletypesize{\scriptsize}
\tablehead{
Observations &
\multicolumn{2}{c}{ Flux density ($\mu$Jy) } \\
  & Reading peak\tablenotemark{a} & Gaussian fittings\tablenotemark{b} \\
} 
\startdata
\multicolumn{3}{c}{JVLA} \\
18--26 GHz (Nov.15) & 114$\pm$17 & 114$\pm$17 \\
18--26 GHz (Nov.17) & 110$\pm$16 & 122$\pm$16 \\
29--37 GHz & 105$\pm$23 & 148$\pm$23 \\
40--48 GHz & 393$\pm$100 & 665$\pm$100 \\\hline
\multicolumn{3}{c}{ALMA} \\
97.5 GHz & \multicolumn{2}{c}{4.1$\pm$0.2\tablenotemark{c}}  \\
145 GHz & \multicolumn{2}{c}{15$\pm$0.75\tablenotemark{c}} \\
%
%
\enddata
\tablenotetext{a}{Flux densities obtained from reading the peak intensities in the clean images.}\vspace{-0.2cm}
\tablenotetext{b}{Flux densities obtained from performing 2D Gaussian fittings in the visibility domain.}\vspace{-0.2cm}
\tablenotetext{c}{Obtained from integrating over the inner 4$''$ square region, nominally assuming a 5\% absolute flux error.}
\end{deluxetable}

\subsubsection{Emission at $\gtrsim$30 GHz}\label{subsub:grown}

The $\gtrsim$30 GHz SED of PDS~70 (Figure \ref{fig:flux}) can be interpreted by dust thermal emission that is optically thick at $\gtrsim$50 GHz and optically thin at lower frequencies.
For example, we compare the observed SED with the SED model that was evaluated based on the DSHARP-default dust opacity table, taking both absorption and scattering opacities into account (for more details see Appendix \ref{appendix:grown}; c.f. \citealt{Birnstiel2018ApJ...869L..45B}).
We assumed a 20 K mean dust temperature, which may be consistent with that in the PDS~70 dust ring (\citealt{Hashimoto2012ApJ...758L..19H,Hashimoto2015ApJ...799...43H}). 

To gauge the maximum dust grain size ($a_{\mbox{\scriptsize max}}$), in the bottom panel of Figure \ref{fig:flux}, we compared the observed spectral indices with the spectral indices of the dust slabs that have $\Sigma_{\mbox{\scriptsize dust}}=$0.01 g\,cm$^{-2}$ dust column densities and $a_{\mbox{\scriptsize max}}=$0.3 mm, 1.0 mm, 2.0 mm, 3.0 mm, 10.0 mm maximum grain sizes. 
At $\sim$40 GHz frequency, which is the optically thin part of the spectrum, the spectral indices of the dust slab models are maximized when the $a_{\mbox{\scriptsize max}}$ value is in between 1 mm and 2 mm, which makes a good comparison with the observed high spectral index at 33--44 GHz (Figure \ref{fig:flux}, bottom).
Using the other $a_{\mbox{\scriptsize max}}$ values which make the spectral index of this grown dust component smaller will yield a too low spectral index in the integrated SED model, after mixing with free-free emission or the emission of the spinning nanometer-sized dust component that has a very low spectral index (more in Section \ref{subsub:small}; Figure \ref{fig:flux}).
Therefore, in our fiducial model which is presented in the upper panel of Figure \ref{fig:flux}, we adopted $a_{\mbox{\scriptsize max}}=$1.5 mm as a working hypothesis.
The $a_{\mbox{\scriptsize max}}$ in our fiducial model should be regarded as a lower limit.
It is plausible to assume a slightly larger $a_{\mbox{\scriptsize max}}$ value.
However, when $a_{\mbox{\scriptsize max}}$ is comparably greater than 1 cm, the lower absorption opacity spectral index ($\beta$) will start to make the 33--44 GHz spectral index lower, making it harder to explain the 33--44 GHz spectral index.

To match the observed $\gtrsim$30 GHz SED, 
in the fiducial model, the  dust column density ($\Sigma_{\mbox{\scriptsize dust}}$) and solid angle ($\Omega$) were chosen to be 5.0 g\,cm$^{-2}$ and 2$\cdot$10$^{-12}$\,Sr, respectively.
The overall dust mass in this grown dust component is $\sim$180 $M_{\oplus}$, which is uncertain due to the assumptions of dust opacities.

The values of $\Sigma_{\mbox{\scriptsize dust}}$ and $\Omega$ are degenerated and thus were loosely constrained. 
Nevertheless, the  $\Omega$ given in our fiducial model should be regarded as an upper limits, and thus the $\Sigma_{\mbox{\scriptsize dust}}$ should be a lower limits.
We found that if we increase $\Omega$ and decrease $\Sigma_{\mbox{\scriptsize dust}}$ by the same factor, the flux densities at 97.5 GHz and 145 GHz will immediately exceed what have been measured in the ALMA observations (Table \ref{tab:flux}).
On the other hand, for a model with a 2 times larger $\Sigma_{\mbox{\scriptsize dust}}$ and a 2 times smaller $\Omega$, the emission at 33 and 44 GHz will be too optically thick to reproduce the observed, 5$\pm$1 spectral index (Figure \ref{fig:flux}).

The solid angle $\Omega$ is considerably smaller than that of the millimeter ring resolved at $>$200 GHz frequencies (\citealt{Benisty2021ApJ...916L...2B,Isella2019ApJ...879L..25I,Casassus2019MNRAS.483.3278C}).
It is comparable to that of a circular dust slab that has a $\sim$0\farcs18 angular radius, while the geometry of the grown dust component was not resolved in our JVLA observations. 
For example, the grown dust component in our model may represent either a few spatially unresolved clumps,  or a crescent that has $\sim$0\farcs1 characteristic width and $\sim$0\farcs3 characteristic length, or a combination of those two cases. 
This is consistent with the shape of the crescent in the northwest (Figure \ref{fig:almaimage}; see also Figure 8 of \citealt{Benisty2021ApJ...916L...2B}).
This crescent may be a local concentration of grown dust, which is similar to what has been recently spatially resolved in the DM~Tau disk (\citealt{Liu2024A&A...685A..18L}).

Finally, we remark that if we extrapolate $F_{\mbox{\scriptsize 97.5 GHz}}$ to 44 GHz by assuming a spectral index of 2.0, the extrapolated 44 GHz flux density will be 840 $\mu$Jy, which is only 26\% higher than our measurement listed in Table \ref{tab:flux}.
This extrapolation implies that our 44 GHz flux density measurement was unlikely to be significantly underestimated, otherwise, the 44--97.5 GHz spectral index will become lower than 2.0, which seems unlikely.
It is less likely that our 44 GHz flux density measurement was overestimated, since the calibration errors tend to yield decoherent signal.
Usually, the JVLA Q band (40--48 GHz) observations are the most difficult to calibrate.
Our measurements at lower frequencies were likely less affected by observational issues or calibration errors.
Since our two epochs of 22 GHz measurements achieved good consistency,
they are unlikely to have been biased by serious calibration errors (Table \ref{tab:flux}).
We do not have any evidence that the 33 GHz flux density was
seriously biased (i.e., underestimated) due to calibration issues.
Even if the 33 GHz flux density was underestimated, the interpretation for the observed SED (Figure \ref{fig:flux}) will be qualitatively unchanged. 
In that case, we will have to use a larger $\Sigma_{\mbox{\scriptsize dust}}$ value for the grown dust component (Appendix \ref{appendix:uncertainty}).

\begin{figure}
    \hspace{-0.6cm}
    \includegraphics[width=9.1cm]{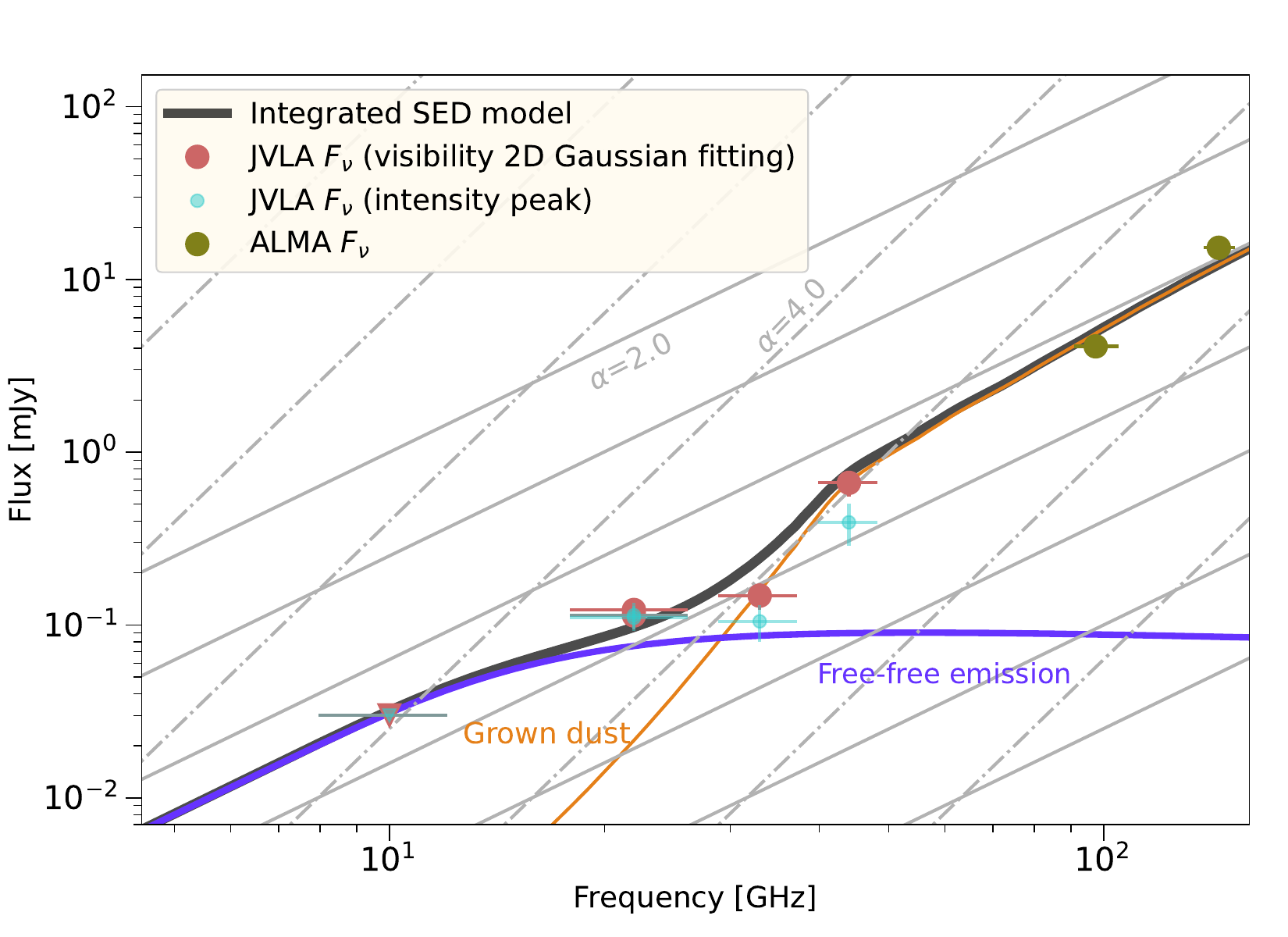}
    \caption{
    Similar to the top panel of Figure \ref{fig:flux}, but use free-free emission (Appendix \ref{appendix:freefree}) instead of the emission of spinning nano-meter sized dust to model the flux densities at low frequencies.
    }
    \vspace{0.37cm}
    \label{fig:fluxfreefree}
\end{figure}

\subsubsection{Emission at $\lesssim$30 GHz}\label{subsub:small}

With an assumed $\sim$20 K dust temperature, the emission at 22--33 GHz has a too low spectral index (0.6$\pm$0.2; Figure \ref{fig:flux}) to be interpreted with the continuum emission of an isothermal dust slab that has a $>$0.1 $\mu$m $a_{\mbox{\scriptsize max}}$, even with the consideration of the anomalous reddening effect (\citealt{Liu2019ApJ...877L..22L}).
In some previous studies on the relatively massive disks, such low spectral indices were interpreted by the emission of optically thick dust sources that has a temperature gradient along the line-of-sight (e.g., \citealt{Li2017ApJ...840...72L,Galvan2018ApJ...868...39G,Liu2019ApJ...884...97L,Wright2022ApJ...924..107W}).
This cannot be the case of PDS~70 since the high spectral index at 33--44 GHz (Figure \ref{fig:flux}; Table \ref{tab:flux}) needs to be interpreted with optically thin dust emission in the Rayleigh-Jeans limit (\citealt{Hildebrand1983QJRAS..24..267H}).

To interpret the SED of PDS~70, besides the thermal emission of the normally considered dust ($a_{\mbox{\scriptsize max}}>$0.1 $\mu$m), another emission mechanism is needed. 
We first discuss the possibility of interpreting the SED with normal grown dust thermal emission (Appendix \ref{appendix:grown}) and free-free emission (Appendix \ref{appendix:freefree}). 
The spectrum of the grown dust component is the same as that of the grown dust model introduced in Section \ref{subsub:grown} (Figure \ref{fig:flux}).

The spectral index of optically thin free-free emission is $\sim-$0.1.
In principle, the spectral index of optically thick free-free emission can be as high as 2.0 (\citealt{Rybicki1986rpa..book.....R}).
However, an entirely optically thick free-free emission source with a $\sim$2.0 spectral index is not necessarily realistic. 
The realistic ionized wind or jet naturally has an optically thick core and the relatively optically thin halo (\citealt{Reynolds1986ApJ...304..713R}). 
Therefore, the spectral index of unresolved, relatively optically thick ionized wind or jet is theoretically expected to be $\sim$0.6 (\citealt{Wright1975MNRAS.170...41W,Angelada1998AJ....116.2953A}), which is considerably lower than our constrained lower limit at 10--22 GHz.
If the actual 10--22 GHz spectral index is constrained to be $>$2, it will become impossible to interpret the 22 GHz emission with free-free emission. 
Moreover, since the optical depths and spectral indices of free-free emission do not vary with frequency abruptly (Appendix \ref{appendix:freefree}), when using optically thick free-free emission to interpret the X and K bands non-detection/observations with a $>$1.7 spectral index, we found that the SED model will inevitably show large excess at Ka band (33 GHz; Figure \ref{fig:fluxfreefree}; see also the discussion in Appendix \ref{appendix:uncertainty}).
Due to these reasons, we disfavor interpreting the $\lesssim$30 GHz emission only with free-free emission.

It is not impossible that the source PDS~70 presented radio variability although the previous radio monitoring surveys have indicated that the chance is low (e.g. \citealt{Liu2014ApJ...780..155L,Dzib2015ApJ...801...91D,Coutens2019A&A...631A..58C,Getman_2022}). 
We note that the hosts of many known disks that presented radio flares are close binaries (\citealt{Salter_2008,Curiel2019ApJ...884...13C,Chung2024arXiv240519867C}) while PDS~70 is not a binary system. 
If PDS~70 underwent a flare of free-free emission during the two epochs of our K band (22 GHz) observations and was quiescent during our X band (10 GHz) observations, the apparent 10--22 GHz spectral index could be biased high. 
Nevertheless, we do not think this is particularly likely, due to the proximity of our X band and K band observations in time (Table \ref{tab:jvla}).
Moreover, we obtained very consistent 22 GHz flux densities in the two epochs of K band observations (Table \ref{fig:flux}), which may disfavor the interpretation with significant and short duration radio variability. 

The 22--44 GHz spectral profile in Figure \ref{fig:flux} appears similar to a combination of the Rayleigh-Jeans tail of the standard thermal emission of grown dust (c.f. \citealt{Hildebrand1983QJRAS..24..267H}) and the anomalous microwave emission (AME; for a review see \citealt{Dickson2018NewAR..80....1D}).
In the following discussion, we consider the possibility of attributing the origin of AME to spinning nanometer-sized dust (\citealt{Draine2012ApJ...757..103D,Hoang2018ApJ...862..116H}).

In Figure \ref{fig:flux}, we compare our JVLA observations with the emission of a spinning nanometer-sized dust component and the emission of the grown dust component (the same as the grown dust component introduced in Section \ref{subsub:grown}).
We approximated the SED of the spinning nanometer-sized dust component by quoting the empirical formula in \citet{Hoang2018ApJ...862..116H} (see Appendix \ref{appendix:spin}).
This SED model mimics the spectral shape of the observational data, however, has a tension with the 3-$\sigma$ upper limit at X band (10 GHz).
This tension is merely due to that the empirical formula has a too low spectral index in the low frequency tail.
This tension can be largely alleviated if the spectral profile of the spinning nano-meter sized dust is narrower.
If we adopt the more realistic SED model of spinning nanometer-sized dust particles (c.f. \citealt{Rafikov2006ApJ...646..288R,Hoang2018ApJ...862..116H}), the narrower spectral profile can be achieved by using a smaller characteristic grain size (e.g. $\lesssim$2\AA) and a small characteristic width of grain size distribution (e.g., $\lesssim$0.2 in the log space)
The detailed model for spinning nano-meter sized dust is beyond the scope of the present observational study.
We remark that if we take our JVLA Q band measurement of the PDS70 flux density as a spurious detection and consider the Q band flux density quoted in Table \ref{tab:flux} as an upper limit
(Section \ref{sec:results}; Figure \ref{fig:image}), then the spinning nanometer-sized dust would be the most prominent emission source at 22 and 33 GHz (see also the discussion in Appendix \ref{appendix:uncertainty}).

\subsection{Physical implication}\label{sub:implication}

In a protoplanetary disk, individual nanometer-sized dust particles may carry net electric charges.
Nanometer-sized dust particles thus play an important role in controlling how the disk is coupled with the magnetic field.
Forming a Keplerian disk may require depleting the nanometer-sized dust particles, otherwise, the magnetic braking can be too efficient (\citealt{Zhao2016MNRAS.460.2050Z}).
In a protoplantary disk, nanometer-sized dust particles may be replenished by destructing larger-sized dust particles (e.g., \citealt{Hoang2019NatAs...3..766H}).

Infrared observations can detect nanometer-sized dust particles at the disk surfaces (e.g., \citealt{Geers2009A&A...495..837G,Kokoulina2021A&A...652A..61K} and references therein), however, do not sample the disk mid-plane. 
Whether or not nanometer-sized dust particles are present in the disk mid-plane has been controversial. 
Some previous observations on low-mass (\citealt{Greaves2018NatAs...2..662G,Hoang2018ApJ...862..116H,Curone2023A&A...677A.118C}) and massive disks (\citealt{Wright2023ApJ...945...14W}) claimed that the  microwave emission of spinning nanometer-sized dust can explained the centimeter SEDs of their target sources. 
Nevertheless, it is not necessarily the only plausible interpretation for their data.
\citet{Wright2023ApJ...945...14W} also considered the possibility of interpreting the same observations with $e^{-}$ free-free emission which is relevant to massive disks, while \citet{Curone2023A&A...677A.118C} preferred an interpretation of strong free-free variability.

The present work reinforces that it is realistic to consider the emission of spinning nanometer-sized dust particles in the modeling of $\lesssim$50 GHz SEDs, although it is not yet clear whether or not this component is prominent in every protoplanetary disk.
The emission of the spinning nanometer-sized dust particles may resemble the 30--50 GHz excess of millimeter-sized dust emission.
The possibility of detecting the emission of spinning nanometer-sized dust particles thus makes it harder to confirm the presence of grown dust if nanometer-sized dust particles commonly present in protoplanetary disks (\citealt{Greaves2022MNRAS.513.3180G}). 
In this case, simultaneously detecting the anomalously reddened and bluened features in the well-sampled SED (\citealt{Liu2019ApJ...877L..22L,Liu2021ApJ...923..270L}) and spatially resolving the polarization of dust self-scattering (\citealt{Kataoka2015ApJ...809...78K}) may be the most robust approach of confirming dust growth in the millimeter-sized regime. 
The wide bandwidth survey (e.g., 1--50 GHz) on a large number of protoplanetary disks with the JVLA, or in the future with the Next Generation Very Large Array (ngVLA) and/or the Square Kilometer Array (SKA), will shed light on these issues.

\section{Conclusion} \label{sec:conclusion}
 We present the first Karl G. Jansky Very Large Array (JVLA) observations on the PDS~70 disk at Q (40--48 GHz), Ka (29--37 GHz), K (18--26 GHz), and X (8--12 GHz) bands that were taken from late 2019 and early 2020, and the complementary ALMA Bands 3 (97.5 GHz) and 4 (145 GHz) observations.
We detected PDS~70 at modest significance at Ka and K bands and at a marginal significance at Q band; we did not detect PDS~70 at X band.
The PDS~70 ring appears lopsided in the ALMA images.
The lopsidedness likely signifies the presence of a dust crescent in the northwest. 
Based on the resolved spectral profile and the X band non-detection, we suggest that the flux density at $\gtrsim$33--44 GHz is dominated by the optically thin thermal emission of a grown dust component that may have a $\gtrsim1$ mm maximum grain size. 
The grown dust component may represent substructures (e.g., the dust crescent) in the PDS~70 ring, given that is solid angle is smaller than that of the bulk of the PDS~70 ring.
At lower frequencies the flux densities are not consistent with the standard dust emission, and the most plausible explanation is that they are dominated by the emission of spinning nanometer-sized dust particles
The future JVLA, ngVLA, and SKA surveys are important to address how common is the emission feature of the nanometer-sized dust particles in the protoplanetary disks. 

\begin{acknowledgments}
We thank the referee for the useful suggestions. 
This paper makes use of the following ALMA data: ADS/JAO.ALMA\#2022.1.01477.S. ALMA is a partnership of ESO (representing its member states), NSF (USA) and NINS (Japan), together with NRC (Canada), NSTC and ASIAA (Taiwan), and KASI (Republic of Korea), in cooperation with the Republic of Chile. The Joint ALMA Observatory is operated by ESO, AUI/NRAO and NAOJ.
The National Radio Astronomy Observatory is a facility of the National Science Foundation operated under cooperative agreement by Associated Universities, Inc.
H.B.L. is supported by the National Science and Technology Council (NSTC) of Taiwan (Grant Nos. 111-2112-M-110-022-MY3).
T.M. is supported by JSPS KAKENHI Grant Number JP23K03463.
K.D. is supported by JSPS KAKENHI Grant Number JP22KJ1435. 
\end{acknowledgments}

%

\facilities{JVLA}


\software{
          astropy \citep{2013A&A...558A..33A,astropy2018AJ....156..123A,astropy2022ApJ...935..167A},  
          Numpy \citep{VanDerWalt2011}, 
          CASA \citep{McMullin2007ASPC..376..127M},
          }

\appendix

\section{Discussion on the impact of flux extraction systematics}\label{appendix:flux}
It is common to measure the overall flux densities by performing 2D Gaussian fittings in the (\texttt{clean}) image domain or by integrating the flux densities over the regions in the \texttt{clean} image where the target source(s) is detected above a certain significance level (e.g., 2-$\sigma$).
These methods work well when the intensity distribution of the target source is resolved with high significance, such that the sum of the flux density in the \texttt{clean} components is close to the overall flux density of the target source.

However, when the intensity distribution is resolved at modest significance (e.g., 2--4-$\sigma$), it is difficult to recover a high fraction of the overall flux densities in the \texttt{clean} components.
Nevertheless, as long as the innermost region of the synthesized dirty beam is approximately Gaussian, the sum of the flux densities in the \texttt{clean} components and the flux densities in the \texttt{clean} residual image is still representative of the overall flux density of the target source (c.f. Appendix of \citealt{Casassus2022MNRAS.513.5790C}).
Therefore, the two methods of measuring flux densities in the image domain mentioned above may still work reasonably well.

When the synthesized dirty beam is highly non-Gaussian, the \texttt{clean} components and the \texttt{clean} residual image are represented by very different beams (i.e., point spread functions), leading to an ambiguous Jy\,beam$^{-1}$ intensity unit in the finally restored \texttt{clean} images.
In such cases, integrating flux densities in the image domain will yield biased overall flux densities.
In the case that the target source is spatially unresolved, and when the residual phase and delay errors are negligible, one can simply measure the overall flux density by reading the peak intensity value (in Jy\,beam$^{-1}$ units) in the image domain, since the Jy\,beam$^{-1}$ unit peak intensity does not depend on the shape of the synthesized dirty and \texttt{clean} beams. 
However, if there are some residual phase and delay errors, the peak intensity should be regarded as a lower limit of the overall flux density since the phase or delay errors can lead to angular dispersion and potential decoherence of the flux densities.

Fitting visibility amplitudes is rather immune to phase errors. 
However, in the visibility domain, it is hard to distinguish any potential confusing sources from the target source. 
Therefore, the flux densities obtained from visibility fittings should be regarded as upper limits.

In each of the two epochs of K band observations, we verified that the following four approaches, (i) performing 2D Gaussian fittings in the \texttt{clean} image domain, (ii) performing 2D Gaussian fittings in the visibility domain, (iii) reading the peak intensities, and (iv) integrating the flux densities over the regions that are above 2-$\sigma$ significance, yielded consistent flux density measurements. 
We found that reading peak intensities in the image domain and performing 2D Gaussian fittings in the visibility domain yield reasonably consistent flux density measurements at Ka and Q bands (Table \ref{tab:flux}).
Making integration or performing 2D Gaussian fittings in the image domain led to significant underestimates of flux densities at Ka and Q bands.

\section{Models of dust emission}\label{appendix:model}

\subsection{Grown dust}\label{appendix:grown}
To produce the SED models for the grown dust component (Figure \ref{fig:flux}), we quoted the DSHARP-default opacity and the Equations (10)--(20) presented  in \citet{Birnstiel2018ApJ...869L..45B}.
For a dust emission source that the angle between the surface normal and the line-of-sight is $\theta$,
those equations evaluate the observed intensity $I^{\mbox{\scriptsize out}}_{\nu}$ at frequency $\nu$ using the Eddington-Barbier approximation 
\begin{equation}
I^{\mbox{\scriptsize out}}_{\nu}\simeq(1-e^{-\Delta\tau/\mu})S_{\nu}\left(  (\frac{1}{2}\Delta\tau-\tau_{\nu})/\mu =2/3\right),
\end{equation}
where $\Delta\tau$ is the total optical depth evaluated based on the effective total extinction, $\tau_{\nu}$ is the optical depth evaluated at a specific scale height, and $\mu=\cos\theta$.
The source function is expressed by
\begin{equation}
S_{\nu}(\tau_{\nu})=\epsilon^{\mbox{\scriptsize eff}}_{\nu}B_{\nu}(T_{\mbox{\scriptsize dust}}) + (1-\epsilon^{\mbox{\scriptsize eff}}_{\nu})J_{\nu}(\tau_{\nu}),
\end{equation}
where $\epsilon^{\mbox{\scriptsize eff}}_{\nu}$ is the effective total extinction at frequency $\nu$, $B_{\nu}(T_{\mbox{\scriptsize dust}})$ is the Planck function at dust temperature $T_{\mbox{\scriptsize dust}}$, and $J_{\nu}(\tau_{\nu})$ is the mean intensity that can be obtained by analytically solving the equations of radiative transfer (\citealt{Miyake1993Icar..106...20M}). 

The dust grains were assumed to have compact morphology  and is composed of water ice \citep{Warren1984ApOpt..23.1206W}, astronomical silicates \citep{Draine2003ARA&A..41..241D}, troilite, and refractory organics (\citealt{Henning1996A&A...311..291H}).
We assumed a power-law grain size distribution function (i.e., $n(a)\propto a^{-q}$) in between the minimum and maximum grain sizes ($a_{\mbox{\scriptsize min}}$, $a_{\mbox{\scriptsize max}}$), where the values of $q$ and $a_{\mbox{\scriptsize min}}$ were assumed to be 3.5 and $10^{-4}$ mm, respectively.
The size-averaged opacities derived based on these assumptions are not sensitive to the assumed  $a_{\mbox{\scriptsize min}}$ values. 


\begin{figure}
    \hspace{-0.6cm}
    \includegraphics[width=9.1cm]{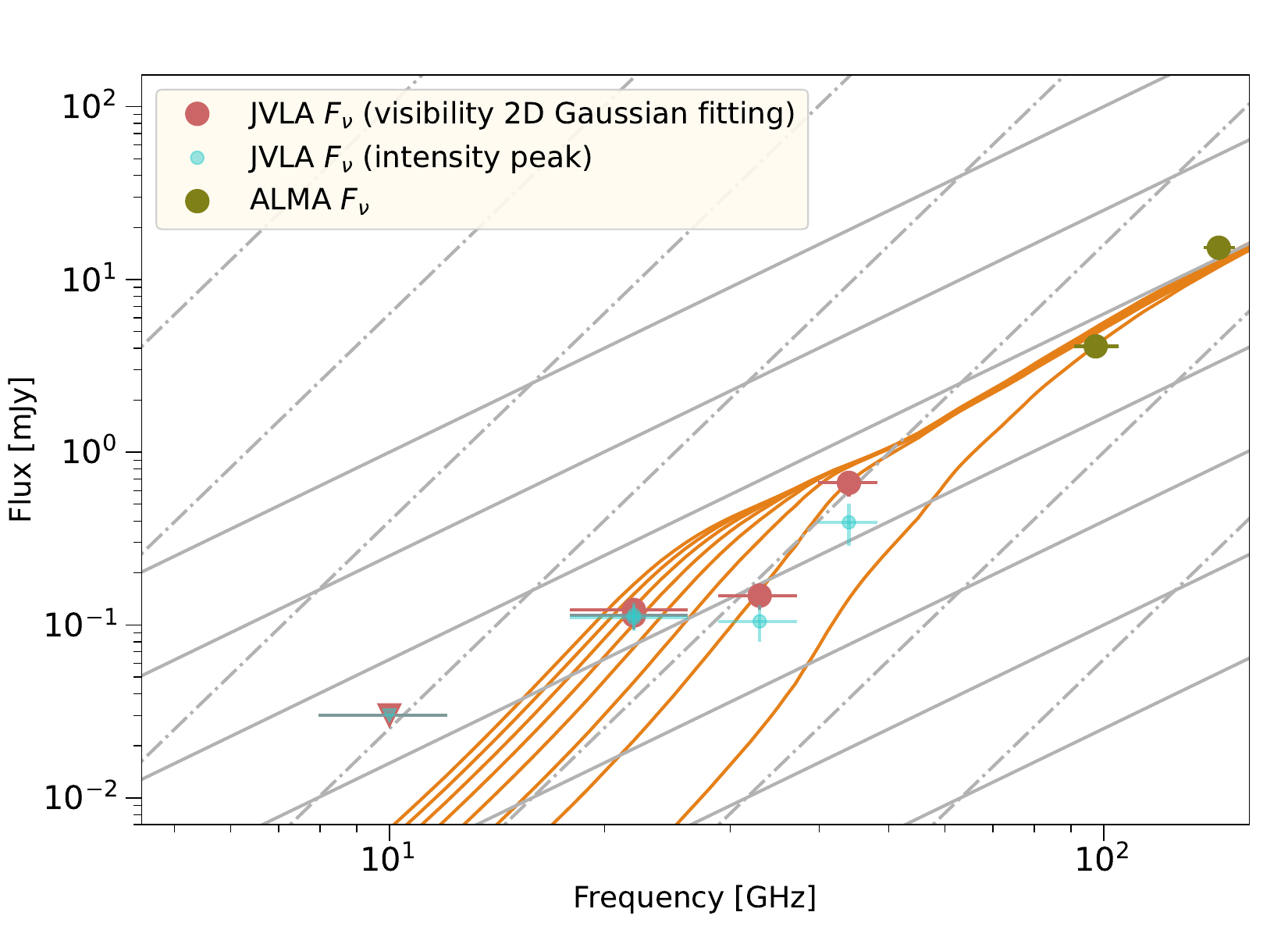}
    \caption{
    Similar to the top panel of Figure \ref{fig:flux}, but only use the thermal emission of a grown ($a_{\mbox{\scriptsize max}}=$1.5 mm) dust component to model the flux densities. 
    There are eight curves, which present the models with $\Sigma_{\mbox{\scriptsize dust}}=$1, 5, 10, 15, 20, 25, 30, 35 g\,cm$^{-2}$, respectively.
    The solid angle and temperature are the same as the assumptions for the grown dust component in Figure \ref{fig:flux}. 
    }
    \vspace{0.37cm}
    \label{fig:fluxdust}
\end{figure}

\subsection{Free-free emission}\label{appendix:freefree}

Free-free emission is the thermal emission of ionized gas that can be described by
\begin{equation}
\label{eq:dust}
F_{\nu} =B_{\nu}(T)(1-{e}^{-\tau_{\nu}^{ff}})\Omega,
\end{equation}
where $\Omega$ is the observed solid angle, $\tau_{\nu}^{ff}$ is the optical depth at frequency $\nu$, and $B_{\nu}(T)$$=$$(2h\nu^{3}/c^{2})$$(e^{h\nu/k_{B}T} -1)^{-1}$ is the Planck function at temperature $T$, $h$ and $k_{B}$ are Planck and Boltzmann constants. 

Following \citet{Keto2003ApJ...599.1196K} and \citet{Mezger1967ApJ...147..471M}, we approximate $\tau_{\nu}^{ff}$ by:
\begin{equation}
\label{eq:tauff}
\tau_{\nu}^{ff}=8.235\times10^{-2}\left(\frac{T_{e}}{\mbox{K}}\right)^{-1.35}\left(\frac{\nu}{\mbox{GHz}}\right)^{-2.1}\left(\frac{\mbox{EM}}{\mbox{pc\,cm$^{-6}$}}\right),
\end{equation}
where $T_{e}$ is the electron temperature, and EM is the emission measure defined as EM$=$$\int n_{e}^{2}d\ell$, $n_{e}$ is the electron number volume density.

For the model present in Figure \ref{fig:fluxfreefree}, we chose $T=8000$ K, EM$=$10$^{9}$ pc\,cm$^{-6}$, and $\Omega=1.3\times10^{-15}$ Sr.
This choice is to make the free-free emission optically thick/thin at below and above $\sim$22 GHz.
These parameters are degenerate due to the lack of observational constraints at $<$22 GHz. 

We note that Equations \ref{eq:dust} and \ref{eq:tauff} describe the thermal emission of an ionized gas slab that has uniform temperature and surface density.
When EM is arbitrarily high, such uniform ionized gas slabs can be optically thick and thus show $\sim$2.0 spectral indices at a specific frequency.
However, in spite that an optically thick ionized gas slab can reproduce the $>$1.7 spectral index in PDS~70 at 10--22 GHz, such ionized gas slab may not represent the realistic ionized gas structures in young stellar objects. 

In young stellar objects, the ionized gas structures that are spatially extented enough to contribute to the observed flux densities, are ionized wind/jet (\citealt{Reynolds1986ApJ...304..713R}).
The realistic ionized gas wind/jet will likely have a surface density gradient. 
In that case, the observations at any frequency will always detect a mixture of optically thick and thin emission which makes the spectral index lower than 2.0
(\citealt{Wright1975MNRAS.170...41W,Reynolds1986ApJ...304..713R,Angelada1998AJ....116.2953A}).

\subsection{Spinning nanometer-sized dust}\label{appendix:spin}
We quote Equation (15) of \citet{Hoang2018ApJ...862..116H} to represent the SED of the spinning nanometer-sized dust particles.
The equation is reproduced as follows:
\begin{equation}
F_{\nu} = F_{sd,0} \left( \frac{\nu}{\nu_{pk}} \right) ^{2}e^{1-(\frac{\nu}{\nu_{pk}})^2},
\end{equation}
where $\nu_{pk}$ and $F_{sd,0}$ are the frequency and the flux density at the peak of the spinning dust spectrum, respectively.
The same equation was also utilized in the studies on the SEDs of external galaxies made by \citet{Draine2012ApJ...757..103D}.

We adopted $\nu_{pk}=$21 GHz and $F_{sd,0}=$0.07 mJy.
We note that this equation is an empirical formulation rather than the full analytical expression for the spectrum of the spinning nano-meter sized dust. 
We argue that quoting this Equation is appropriate for our present purpose of comparing with the observations that were not very well sampled in the frequency domain. 
The more sophisticated SED model of spinning
nanometer-sized dust particles, which involves a large number of free parameters (\citealt{Rafikov2006ApJ...646..288R,Hoang2018ApJ...862..116H}), may be tested after the detailed spectral profile at $\sim$5--50 GHz is resolved by future observations.

\subsection{Consideration of measurement uncertainties}\label{appendix:uncertainty}

In Section \ref{sec:discussion}, we discussed the SED model that fits the JVLA flux density measurements made by fitting 2D Gaussian in the visibility domain (Table \ref{tab:flux}).
Here we briefly discuss the dust emission model that fits the JVLA flux density measurements made by reading the intensity peaks (i.e., the cyan points in Figure \ref{fig:flux}), which we consider have larger uncertainties (Section \ref{sec:results}, Appendix \ref{appendix:flux}).
The two methods of measuring flux densities yielded very consistent measurements at K band (22 GHz) while reading intensity peaks yielded considerably lower flux density measurements at Ka (33 GHz) and Q (44 GHz) bands.

There needs at least one emission mechanism that contributes significantly at K band (22 GHz), which may be either the emission of spinning nanometer-sized dust or free-free emission. 
In either case, fractionally, this emission mechanism contributes more at Ka band if we adopt the intensity peak instead of the results of 2D Gaussian fits.
This makes it harder to explain the observed steep spectral slope at 33--44 GHz (Figure \ref{fig:flux}).

When we adopt the intensity peak instead of the results of 2D Gaussian fits, free-free emission cannot be a prominent emission mechanism at K band as it would contribute too much emission at low spectral index at 33--44 GHz frequencies.
The emission of spinning nanometer-sized dust may be the only plausible dominant emission mechanism at K band while there needs some adjustments in the grain size distributions to make the flux density drops rapidly at $>$26 GHz frequencies  (c.f. \citealt{Hoang2018ApJ...862..116H}).

Finally, if we ignore the 33 GHz measurement (e.g., considering that it is subject to a large systematic calibration error), it would be possible to fit the 22 GHz and 44 GHz measurements by only considering a grown dust component ($a_{\mbox{\scriptsize max}}\gtrsim$1 mm), without considering the emission of spinning nanometer-sized dust (Appendix \ref{appendix:spin}).
However, in this case, the close to $\sim$2.0 spectral index between 22 GHz and 44 GHz (Figure \ref{fig:flux}) will imply that the dust emission is not optically thin even at $\sim$22 GHz.
To achieve such a high optical depth, the lower limit of dust mass will be $\sim$5 times higher than the dust mass given by our fiducial model, which appears unrealistic to us (Figure \ref{fig:fluxdust}). 
In this sense, it is realistic to consider that at least part of the 22 GHz flux density is contributed by the emission of spinning nanometer-sized dust.




\bibliography{main}{}
\bibliographystyle{aasjournal}
\end{document}